%% file: 00_MAIN.tex
\documentclass{aastex631}

\usepackage{amsmath}	
\usepackage{amssymb}	
\usepackage[english]{babel}
\usepackage{blindtext}  
\usepackage{booktabs}   
\usepackage{graphicx}	
\usepackage{mhchem}     
\usepackage{multirow}   
\usepackage{pifont}     
\usepackage[figuresleft]{rotating}

\shorttitle{Reduced Post-Impact Atmospheres}

\graphicspath{{./}{figures/}}

\begin{document}

\title{Reduced atmospheres of post-impact worlds: The early Earth}

\author[0000-0003-2079-8171]{Jonathan P. Itcovitz}
\affiliation{Institute of Astronomy, University of Cambridge, Madingley Road, Cambridge, CB3 0HA, UK}
\email{ji263@cam.ac.uk}

\author{Auriol S.P. Rae}
\affiliation{Department of Earth Sciences, University of Cambridge, Downing Street, Cambridge CB2 3EQ, UK}

\author{Robert I. Citron}
\affiliation{Department of Earth and Planetary Sciences, University of California, Davis, CA95616, USA}

\author{Sarah T. Stewart}
\affiliation{Department of Earth and Planetary Sciences, University of California, Davis, CA95616, USA}

\author{Catriona A. Sinclair}
\affiliation{Institute of Astronomy, University of Cambridge, Madingley Road, Cambridge, CB3 0HA, UK}

\author{Paul B. Rimmer}
\affiliation{Cavendish Astrophysics, University of Cambridge, JJ Thomson Avenue, Cambridge CB3 0HE, UK}
\affiliation{MRC Laboratory of Molecular Biology, Francis Crick Avenue, Cambridge CB2 0QH, UK}

\author{Oliver Shorttle}
\affiliation{Institute of Astronomy, University of Cambridge, Madingley Road, Cambridge, CB3 0HA, UK}
\affiliation{Department of Earth Sciences, University of Cambridge, Downing Street, Cambridge CB2 3EQ, UK}

\begin{abstract}
Impacts may have had a significant effect on the atmospheric chemistry of the early Earth. Reduced phases in the impactor (e.g., metallic iron) can reduce the planet's \ce{H2O} inventory to produce massive atmospheres rich in \ce{H2}. Whilst previous studies have focused on the interactions between the impactor and atmosphere in such scenarios, we investigate two further effects, 1) the distribution of the impactor's iron inventory during impact between the target interior, target atmosphere, and escaping the target, and 2) interactions between the post-impact atmosphere and the impact-generated melt phase. We find that these two effects can potentially counterbalance each other, with the melt-atmosphere interactions acting to restore reducing power to the atmosphere that was initially accreted by the melt phase. For a $\sim10^{22}\,\mathrm{kg}$ impactor, when the iron accreted by the melt phase is fully available to reduce this melt, we find an equilibrium atmosphere with \ce{H2} column density $\sim10^4\,\mathrm{moles\,cm^{-2}}$ ($p\mathrm{\ce{H2}}\sim120\,\mathrm{bars}\mathrm{,}~X_\mathrm{H2}\sim0.77$), consistent with previous estimates. However, when the iron is not available to reduce the melt (e.g., sinking out in large diameter blobs), we find significantly less \ce{H2} ($7\times10^2-5\times10^3\,\mathrm{moles~cm^{-2}}\mathrm{,}~p\mathrm{\ce{H2}}\lesssim60\,\mathrm{bars}\mathrm{,}~X_\mathrm{H2}\lesssim0.41$). These lower \ce{H2} abundances are sufficiently high that species important to prebiotic chemistry can form (e.g., \ce{NH3}, \ce{HCN}), but sufficiently low that the greenhouse heating effects associated with highly reducing atmospheres, which are problematic to such chemistry, are suppressed. The manner in which iron is accreted by the impact-generated melt phase is critical in determining the reducing power of the atmosphere and re-solidified melt pool in the aftermath of impact.
\end{abstract}



\input{sections/1_introduction}
\input{sections/2_pre_impact}
\input{sections/3_impact_processing}

\input{sections/4_equilibration}
\input{sections/5_results}
\input{sections/6_discussion}
\input{sections/7_conclusion}

\section*{}
\begin{center}
    The code used for the calculations carried out in this study, as well as for the creation and replication of Figures \ref{fig:GADGET Melt Masses}, \ref{fig:Iron Distribution}, \ref{fig:Init Atmos}, \ref{fig:Walk-Through}, \ref{fig:Five Models}, and \ref{fig:Mantle Remixing} is available at \url{https://github.com/itcojo/itcovitz_reduced_atmospheres}.
    \vspace{3mm}
\end{center}

\bibliographystyle{aasjournal}
\bibliography{references/refs} 

\begin{acknowledgments}
With thanks to the anonymous reviewers for helpful comments. The authors would like to thank Simon Lock for his role in helping to bring us together, and Gareth Collins for useful conversations. The authors acknowledge \citet{wong2011color} for enabling the creation of accessible graphics. This work was supported by the UK Science and Technology Facilities Council (STFC) grant number ST/T505985/1. A.S.P.R. gratefully acknowledges funding from Trinity College Cambridge. R.I.C. and S.T.S. are supported by the Simons Collaboration on the Origins of Life grant number Simons-554203.
\end{acknowledgments}

\clearpage
\appendix
\input{appendices/A_GADGET}

\input{appendices/B_simultaneous}
\input{appendices/C_post_impact}

\end{document}

%% file: sections/1_introduction.tex
\section{Introduction} \label{sec:Introduction}
Impacts have the potential to produce substantial changes in the atmospheres of young terrestrial planets. The energy added to the planet, as well as compositional differences between the accreted impactor materials and the regions of the planet which they accrete to, can lead to large-scale thermochemical alterations to the planet's surface, atmosphere, and interior. Such alterations can be long-lived (e.g., the Moon forming impact, \citealt{canup2012forming, cuk2012making, sleep2014terrestrial, lock2018origin}) or short-lived (e.g., reducing surface environments,  \citealt{benner2020when, zahnle2020creation}).

Reduced geochemical environments are of particular interest because they are required for many prebiotic chemical pathways. These pathways generate RNA precursors starting from reduced species such as \ce{CH4}, \ce{NH3}, \ce{PH3}, and \ce{HCN} \citep{oro1961mechanism, sutherland2016origin}. Reduced environments can host such atmospheric species over extended timescales, and can make them readily available at the planet surface through rainout \citep{benner2020when}. 
\begin{figure*}
    \centering
    \includegraphics[width=0.98\textwidth]{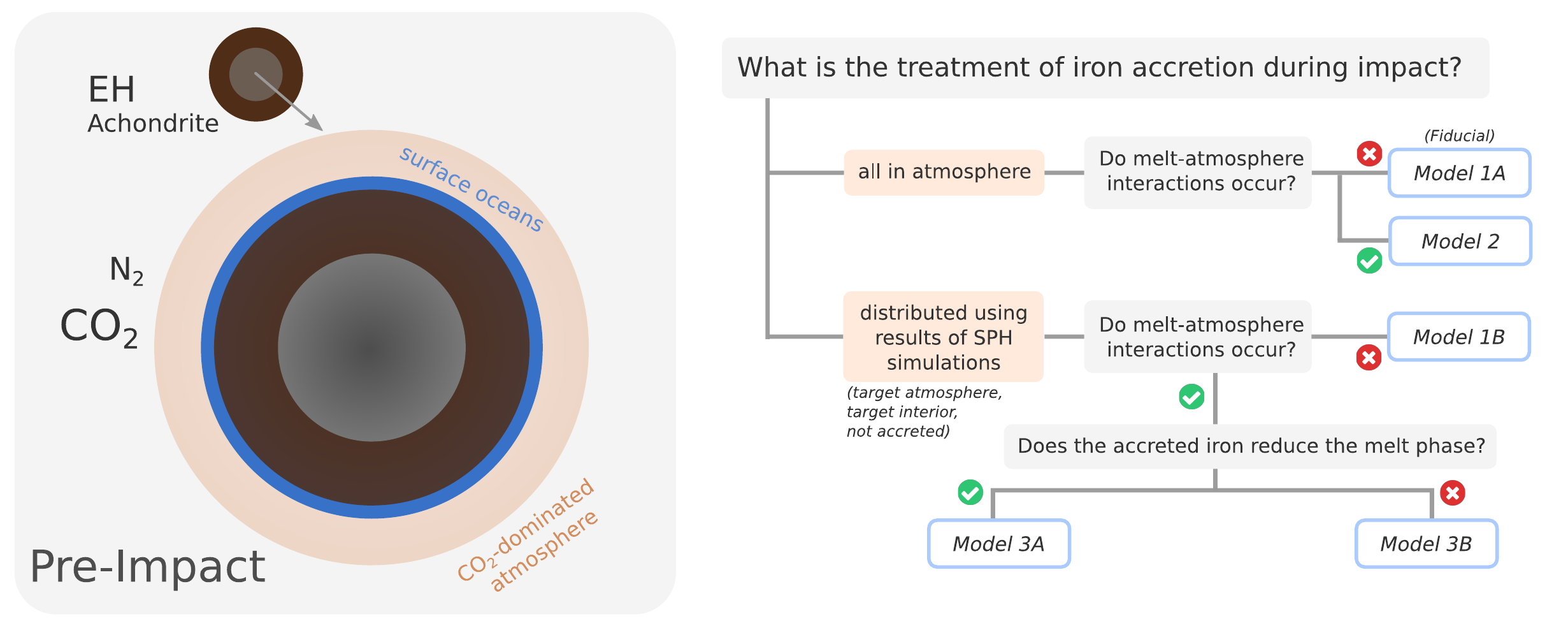}
    \caption{\textit{Left:} Vertical slice of the target and impactor (impact angle of $45^\circ$) before the impact takes place, showing the mantles and cores of both bodies, as well as the target's surface conditions, \textit{Right:} decision tree showing the components of the physical model included in each of the five models presented in Section \ref{sec:Results}.}
    \label{fig:Graphical Abstract}
\end{figure*}

The compositions of terrestrial planet atmospheres are, however, broadly oxidised, making them unsuitable for prebiotic chemistry. The primordial atmospheres which rocky planets accrete from their protoplanetary discs during formation are commonly lost (e.g., \citealt{ginzburg2016super, owen2017evaporation}), if they indeed are ever accreted substantially, and secondary atmospheres are outgassed from their interiors (e.g., \citealt{gaillard2014theoretical, liggins2020can}). The composition of a planet's mantle will thus determine the chemistry of the its long-term secondary atmosphere. In the case of Earth, its ancient mantle is suggested to have had an oxygen fugacity ($f$O$_2$) near the relatively oxidised Fayalite-Magnetite-Quartz (FMQ) buffer \citep{trail2011oxidation}. Earth's early outgassed atmosphere would thus have been rich in \ce{CO2} and \ce{H2O}, but relatively poor in \ce{H2} and the species suggested to be important for prebiotic chemistry (e.g., \citealt{liggins2020can}). Such arguments suggest that reduced atmospheres are unlikely to be common on rocky planets over geological timescales.

While oxidised atmosphere may be problematic for initiating prebiotic chemistry, having globally reduced environments could itself be problematic for sustaining habitability in the long term. One reason is that, due to abundant greenhouse gases (e.g., \ce{H2}, \ce{CH4}), these environments can host surface temperatures well above those suitable for prebiotic chemistry \citep{pierrehumbert2011hydrogen, wordsworth2013hydrogen}. Specific prebiotic chemical scenarios may also face challenges in reduced atmospheres. For example, the UV light necessary in the chemistry of \citet{powner2009synthesis} could be blocked by the organic hazes that can form in such environments, although hazes can be somewhat suppressed under extremely reducing conditions \citep{trainer2006organic}.

The most favourable global environments for prebiotic chemistry may, therefore, be transient. Here, an oxidising surface environment experiences a perturbation that makes it temporarily reducing. Reduced species form in the environment, laying the foundations for prebiotic pathways that continue while the the planet relaxes back to its oxidised state \citep{benner2020when}.

Large impacts (i.e., Ceres- to Pluto-sized projectiles) have become strong candidates for producing such global transient reduced environments through their ability to generate large abundances of \ce{H2}. The \ce{H2} is produced through evaporation of the planet's surface \ce{H2O} inventory during impact and reduction of this \ce{H2O} by metallic iron from the impactor core \citep{genda2017terrestrial}. The massive steam atmosphere then radiatively cools (e.g., \citealt{abe1988evolution, kasting1988runaway, zahnle1988evolution}), favouring the production of \ce{CH4} and \ce{NH3} through thermochemical reactions of the impact-generated \ce{H2} with the planet's initial atmospheric C- and N-bearing molecules (Figure \ref{fig:Graphical Abstract}, \textit{left}). The atmosphere eventually cools sufficiently such that the steam rains out from the atmosphere and reforms oceans on the planet surface. The remaining atmosphere is reducing in nature. \ce{HCN} can then form via photochemistry \citep{abelson1966chemical, zahnle1986photochemistry}, initiating a suite of prebiotic chemical pathways \citep{powner2009synthesis, ritson2012prebiotic, patel2015common}.

Evidence for large impacts in Earth's history after Moon formation is indirectly visible in the Earth's mantle, where highly siderophile elements (HSEs) are present in excess of formation models' predictions \citep{mann2012partitioning, rubie2015accretion}. These HSEs have strong tendencies to be incorporated into the core alongside iron during planet formation. Mantle excesses, and particularly the elemental proportions in which we find them, thus indicate accretion of chondritic material after core formation had ended \citep{rubie2015accretion, rubie2016highly}. Comparison between Earth's HSE excesses and those in the lunar mantle suggests that the Earth accreted a greater total mass through these late impacts than the Moon, even accounting for their relative gravitational cross sections \citep{day2016highly}. A stochastic accretion model can explain such measurements with a single large impactor of approximately chondritic composition \citep{bottke2010stochastic, brasser2016late, day2016highly}. However, lower retention rates of late accretion material by the Moon in comparison to the Earth have also been suggested as possible explanations for this discrepancy in HSEs \citep{kraus2015impact, zhu2019reconstructing}.
\begin{table}
    \caption{Model parameters used as our set of standard values.}
    \label{tab:Standard Parameters}
    \parbox{.45\linewidth}{
    \begin{tabular}{rcll}
        \toprule
        Property                 &  Symbol            &  Value               &  Units              \\
        \midrule
        Target Mass              &  $M_t$             &  $5.9\times10^{24}$ &  kg                 \\
        \quad                    &  \quad             &  1.0                &  M$_\mathrm{E}$     \\
        \midrule
        Atmospheric Composition  &  $p\ce{CO2}$       &  100.0               &  bar                \\ 
        \quad                    &  $p\ce{N2}$        &  2.0                 &  bar                \\
        \midrule 
        Surface Water Inventory  &  $N_\text{ocean}$  &  1.85                &  EO                 \\
        \midrule
        Mantle Water Content    &  $X_\text{H2O}$     &  0.05                &  wt\%               \\
        \midrule
        Melt Ferric Iron Content &  \ce{Fe^3+ / \Sigma\,Fe}  &  \quad        &  \quad              \\
        (Peridotite)             &  \quad                    &  5.0          &  \%                 \\ 
        (Basalt)                 &  \quad                    &  16.0         &  \%                 \\ 
        \midrule
        Melt Oxygen Fugacity     &  $f$O$_2$                 &  \quad        &  \quad              \\
        (Peridotite)             &  \quad                    &   -2.3        &  $\Delta\text{FMQ}$ \\
        (Basalt)                 &  \quad                    &   +0.0        &  $\Delta\text{FMQ}$ \\
        \bottomrule
    \end{tabular}
    }
    ~~~~~~~~~~
    \parbox{.45\linewidth}{
    \vspace{-3.13cm}
    \begin{tabular}{rcll}
        \toprule
        Property                 &  Symbol            &  Value               &  Units              \\
        \midrule
        Impactor                 &  $X_\text{Fe}$     &  33.3                &  wt\%               \\ 
        Composition              &  $X_\text{rock}$   &  66.6                &  wt\%               \\
        \midrule 
        Impact Velocity          &  $v_i$             &  20.7                &  km~s$^{-1}$        \\
        \quad                    &  \quad             &  2.0                 &  $v_\text{esc}$     \\
        \midrule
        Impactor Mass            &  $M_i$             &  $2.0\times10^{22}$  &  kg                 \\
        \midrule
        Impact Angle             &  $\theta_i$          &  $45^\circ$          &  km~s$^{-1}$        \\
        \bottomrule
    \end{tabular}
    }
\end{table}

Previous studies have calculated atmospheric compositions under different large impact scenarios (e.g., \citealt{abe1988evolution, zahnle1988evolution, genda2017terrestrial}), and have also considered such calculations in the context of reduced atmospheres for prebiotic chemistry on early Earth (e.g., \citealt{benner2020when, zahnle2020creation}). However, the main focal point of such studies has been the atmospheric evolution. The influence of how the impactor iron is accreted by the target, and the influence of the planet's impact-generated melt phase, have not been focused upon.

Without models to inform a distribution of the impactor iron inventory during impact, \citet{zahnle2020creation} made the full inventory available to interact with the atmosphere. However, distribution of the iron dictates where in the target the body the reducing power is accreted to (e.g., atmosphere, interior) and how much iron escapes accretion by the target \citep{genda2017terrestrial, marchi2018heterogeneous, citron2022large}. The requisite impactor mass and impact geometry to create a given \ce{H2} inventory is thus unknown without consideration of the iron distribution. These are factors that have important implications for the probability of the impact event occurring, and the compositional fingerprint left behind in the planet's mantle.

Large impacts have the potential to melt significant fractions of the planet's mantle \citep{tonks1993magma, pierazzo1997reevaluation}. Chemical interactions and volatile exchange will take place between the atmosphere and the melt during the period in which the atmospheric composition is dictated by thermochemistry, before the atmosphere cools and the chemistry is kinetically quenched (see Section \ref{sec:Impact Processing}). The melt phase will thus buffer the atmospheric composition and redox state during an important period in the system's evolution after impact. 

This work examines the effect that the distribution of impactor iron, and the interactions between the atmosphere and melt, have on the target planet's post-impact state. We examine these effects over a range of impactor masses and target initial conditions. We do not carry out detailed calculations of atmospheric cooling in the millennia after impact and the associated thermochemistry in the atmosphere, as in \citet{zahnle2020creation}. Our analysis focuses on the redox state of the atmosphere and melt phase, and the \ce{H2} abundance in the atmosphere, at equilibrium (i.e., time-independent; Section \ref{sec:Processing Timeline}). Our calculations can thus be seen as describing the state of the atmosphere before subsequent cooling takes place.

We first transition the system from its pre-impact state (Figure \ref{fig:Graphical Abstract}, \textit{left}) through impact processing to its post-impact state (Figure \ref{fig:Cartoon}), including the distribution of the impactor's iron inventory between the target atmosphere, target interior, and escaping the target. We then calculate the equilibrium state of the combined melt-atmosphere system using a time-independent model. Our calculations involve chemical reactions between the atmosphere and impact-generated melt phase in an \ce{H2}-\ce{H2O}-\ce{Fe2O3}-\ce{FeO}-\ce{Fe} system, as well as the partitioning of \ce{H2O} between the atmosphere and melt. For a more complete description of the timeline of events during and after the impact, the timescales of the processes involved, and hence the justification of using a time-independent model, see Section \ref{sec:Impact Processing}.

We explore several model versions (Figure \ref{fig:Graphical Abstract}, \textit{right}). Each version includes different components of the physical model in order to isolate the key processes involved in establishing our final equilibrium state. In Model 1A (henceforth the Fiducial Model), all of the impactor iron is made available to reduce the atmosphere, and there is no equilibration with the impact-generated melt phase. This is representative of the system at the start of calculations in \citet{zahnle2020creation}, a comparison we justify in Section \ref{sec:Results}. Models 1B and 2 include only one of either the iron distribution or melt-atmosphere interactions, respectively, in order to highlight the effect of each in isolation. In Model 3A, the impactor iron is distributed, and we equilibrate the melt phase and atmosphere. All of the iron accreted by the target interior is made available to reduce the melt phase. This is different in Model 3B, where iron distribution and melt-atmosphere equilibration also take place, but none of the iron accreted by the target interior is made available to reduce the melt phase. Models 3A and 3B are thus end-member cases, representing scenarios where the greatest and least fractions of the impactor iron are accreted by the melt-atmosphere system. We are not able to resolve which model is more accurate under our current suite of simulations, and hence present both cases.

In Section \ref{sec:Pre Impact}, we discuss the ranges of impactor and target properties that we consider. In Section \ref{sec:Impact Processing}, we discuss processing of the atmosphere as a result of the impact, the formation of the impact-generated melt phase, and the distribution of the impactor iron within the target. Section \ref{sec:Equilibration} describes how we then solve for the equilibrium state of the interacting atmosphere and melt phase. Sections \ref{sec:Results} \& \ref{sec:Discussion} present results for each model, interpret their differences, and discuss their implications and limitations.

%% file: sections/2_pre_impact.tex
\section{Pre-impact conditions}\label{sec:Pre Impact}
\subsection{Target properties}\label{sec:Target}
We base the target on early Earth. The target's pre-impact state thus derives from the expected properties of Earth in the first few hundred Myr after the Moon-forming impact (i.e., $\sim4.5-4.3\,\mathrm{Ga}$). We term these the standard values (Table \ref{tab:Standard Parameters}). We would expect somewhat similar properties for Earth-like planets in general at this point in their evolutions. However, variations on these values are also expected, and some remain poorly constrained for Earth itself.

Large impacts onto Earth during the period of late accretion\footnote{Impacts in the period after the Moon-forming impact are often referred to in the context of a Late Heavy Bombardment, although this labelling is now debated and the terms Late Accretion or Late Veneer are preferred \citep{brasser2016late, morbidelli2018timeline}.} are likely to have encountered a relatively cool and wet planet with a \ce{CO2}-dominated atmosphere. At the end of magma ocean solidification, which for Earth is that stemming from the Moon-forming impact, the atmosphere is dominated by gaseous \ce{H2O} and \ce{CO2}  \citep{elkins2008linked, lebrun2013thermal, nikolaou2019factors, catling2020archean}. The planet then cools, leading to the water vapour condensing out into surface oceans. A \ce{CO2}-dominated atmosphere remains, with a significant \ce{N2} component also. The timescale over which the full cycle of crystallisation, cooling, and ocean condensation occurs can vary greatly, depending on how efficiently the \ce{H2O}-\ce{CO2} greenhouse atmospheres can sustain high surface temperatures \citep{abe1988evolution, kasting1988runaway, zahnle1988evolution, hamano2013emergence}. However, such timescales ($\sim\,\mathrm{Myr}$) are likely shorter than the intervals between large impacts \citep{lebrun2013thermal}. Similar initial conditions for the target are, therefore, likely to have been encountered from one impact to the next during late accretion, even if previous large impacts have occurred. 

We take our target surface water inventory as 500 bars in our standard values, akin to 1.85 present day Earth Oceans ($1\,\mathrm{EO}\,\approx 1.37\times10^{21}\,\mathrm{kg}$). $500\,\mathrm{bars}$ is approximately Earth's total (oceans plus mantle) water inventory today \citep{zahnle2007emergence}. We put this full inventory into the surface oceans due to the predicted dryness of the early mantle in the aftermath of magma ocean solidification \citep{abe2000water}. However, based on the timing of water delivery to Earth relative to the epoch of large impacts, as well as the extent of water loss through UV-driven escape, the \ce{H2O} inventory could be lower or higher than 1.85 EO.

We use $100\,\mathrm{bars}$ of \ce{CO2} in our standard initial atmosphere. The suggested abundance of \ce{CO2} in the atmosphere during the period of late accretion varies between several bars and several hundred bars, based on arguments surrounding minimum greenhouse climates and degassing from the preceding magma ocean (e.g., \citealt{sleep2010hadean, zahnle2010earth, catling2020archean}, and references therein). We find that variation of \ce{CO2} in the initial atmosphere between $10-200\,\mathrm{bars}$ has less than a $10\%$ effect on the equilibrium \ce{H2} abundance in the resultant post-impact atmosphere. Hence, while variation in initial \ce{CO2} levels would have consequences for the planet in terms of the thermal structure of the pre-impact atmosphere and surface, we suggest that this would not be consequential during the energetic processing of the planet during impact (Section \ref{sec:Impact Processing}) nor in the resulting steam-dominated atmosphere, and we use a nominal value of $100\,\mathrm{bars}$.
\begin{figure*}
    \centering
    \includegraphics[width=0.94\textwidth]{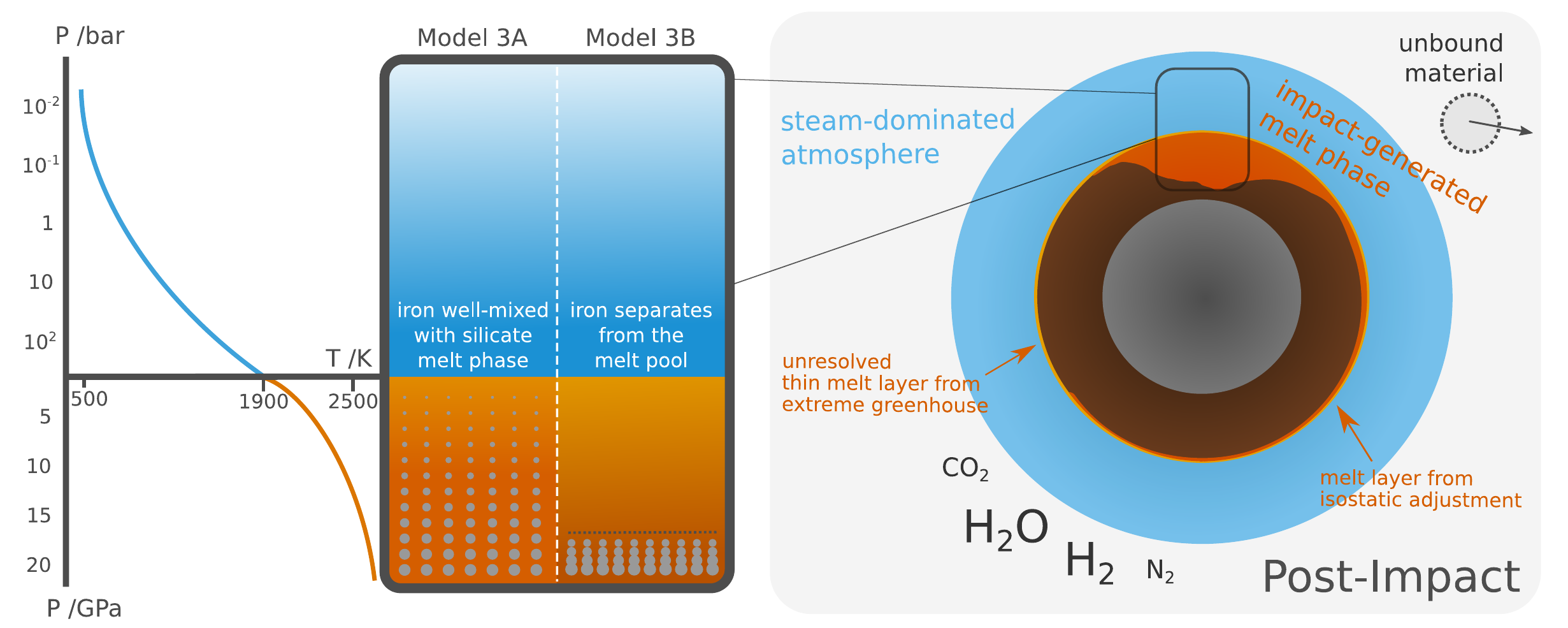}
    \caption{\textit{Left:} Simplified box model used in melt-atmosphere calculations (see Section \ref{sec:Equilibration}). In Model 3A, we assume that the iron accreted by the target interior is well-mixed in the impact-generated melt phase and can chemically interact with the melt. In Model 3B, we assume that this iron instead sinks to the bottom of the melt, and is unable to chemically interact with the melt. An approximate pressure-temperature profile is given, showing the $1900\,\mathrm{K}$ we use as the temperature at the interface between the atmosphere and melt boxes in our model, and which is based off of \citet{kasting1988runaway} and \citet{solomatov1993nonfractional}, \textit{Right:} Vertical slice of the target at the post-impact stage, depicting the settled out structure of the impact-generated melt phase (i.e., the summation of shock melting, adiabatic melting, and isostatic adjustment), the large steam-dominated atmosphere, and material that is gravitationally unbound from planet that we classify as unaccreted by the target.}
    \label{fig:Cartoon}
\end{figure*}

We use $2\,\mathrm{bars}$ of \ce{N2} in our standard initial atmosphere. Earth's present day atmospheric \ce{N2} abundance may be diminished compared to in the Hadean, with an inventory of nitrogen stored in the crust and mantle \citep{johnson2015nitrogen, wordsworth2016atmospheric}. Variation of this value within reasonable bounds has no noticeable effect on our results but may be of importance to the chemical evolution of the system during atmospheric cooling and chemistry in the time after impact, as nitrogen is essential for the formation of \ce{HCN} and \ce{NH3}, key feedstock molecules for prebiotic chemistry.  

We perform calculations with melt phases of both basaltic and peridotitic composition. Both of these melt compositions can form during melting of material of bulk silicate Earth (BSE) composition depending on the degree of melting \citep{kushiro1963origin}. Our simulations suggest that the impact-generated melt phases will be dominated by peridotitic-like melts due to high melt fractions (see Section \ref{sec:Generating Melt}). In these cases, we generate our melt phases with a ferric iron content (expressed as the ratio \ce{Fe^3+ / \Sigma\,Fe}) of $5\,\%$ \citep{davis2021partitioning}, leading to a melt phase $f$O$_2$ around 2.3 log units below the FMQ buffer at post-impact surface temperatures and pressures ($\sim1900\,\mathrm{K}$ and $\sim500\,\mathrm{bars}$). In cases where we use a basaltic melt, representative of low melt fractions (or melting of a thick primary basaltic crust), we generate our melt phases with oxygen fugacity at the FMQ buffer, with $\sim16\,\%$ ferric iron content \citep{cottrell2011oxidation}.

\subsection{Impactor properties} \label{sec:Impactor}
We consider a range of impactor masses. Earth's mantle HSE excesses, if produced by an impactor of roughly chondritic composition, imply a cumulative impact mass of $\sim 2.0\times10^{22}\,\mathrm{kg}$ \citep{bottke2010stochastic}. If we assume that the HSEs were accreted exclusively and efficiently into the mantle by a single event, this mass would be the upper limit on our impactor mass range. However, we show in Section \ref{sec:Iron} that not all of the impactor core accretes during impact. Accounting for this non-accretion, the upper limit on our impactor mass range becomes $\sim 2.4\times10^{22}\,\mathrm{kg}$. The HSE excesses may also have been delivered by several slightly smaller (but still large) bodies in succession. Hence, we consider impactors down to $\sim10\times$ less massive than the maximum HSE impactor. Impactors beneath this range of masses do not see much iron escaping the target (i.e., all iron is usually accreted) and do not produce substantial melt volumes. Hence, the effects demonstrated in this study would have little consequence on our understanding these impacts' perturbation of planetary environments, with results converging on those of previous studies (e.g., \citealt{zahnle2020creation}). However, we emphasise that these smaller impacts are still relevant for producing reduced species in the context of prebiotic chemistry on early Earth. We use a standard impactor mass of $2.0\times10^{22}\,\mathrm{kg}$, as this accounts for a large fraction of the Earth mantle HSE excesses without precluding the earlier or later accretion of some material.

We model all impactors as having the same composition. We assume that, due to their size, the impactors are differentiated \citep{moskovitz2011differentiation, rubie2011heterogeneous}, and separated into an iron-rich core and a silicate mantle. We take their bulk compositions as being close to enstatite (EH) chondrites, as suggested by \citet{dauphas2017isotopic} for bodies striking Earth in the period after Lunar formation. Impactors thus have compositions of $\sim1/3$ metallic iron by mass \citep{mason1966enstatite, keil1968mineralogical, wasson1988compositions} making up the impactor core, and silicate mantle making up $\sim2/3$ by mass. More oxidised compositions of material during late accretion could be possible, such as material similar to carbonaceous chondrites \citep{fischer2017ruthenium, fischer2020ruthenium}. However, reducing conditions would not follow from such impacts, and so we do not consider them here.

We consider a range of impact velocities and impact angles in our simulations of impact melting and iron distribution. A useful unit for expressing the impact velocities is the mutual escape velocity of the target and impactor,
\begin{equation}
    v_\text{esc} = \sqrt{2 G \frac{M_i + M_t}{R_i + R_t}}~~,
    \label{eqn:Mutual Escape Velocity}
\end{equation}
where $M$ represents mass and $R$ represents radius, with subscripts $i$ and $t$ denoting the impactor and target respectively. $G$ is the gravitational constant. In this study, we use expected values as our standard values for both impact velocity and angle. The expected impact velocity for large impacts in the time after Moon formation is $2~v_\text{esc}$ \citep{brasser2020impact}. The most frequent impact angle for an isotropic source of impactors is $45^\circ$ \citep{shoemaker1962interpretation}.

%% file: sections/3_impact_processing.tex
\section{Impact processing of the system} \label{sec:Impact Processing}
\begin{sidewaysfigure}
    \centering
    \includegraphics[width=\textwidth]{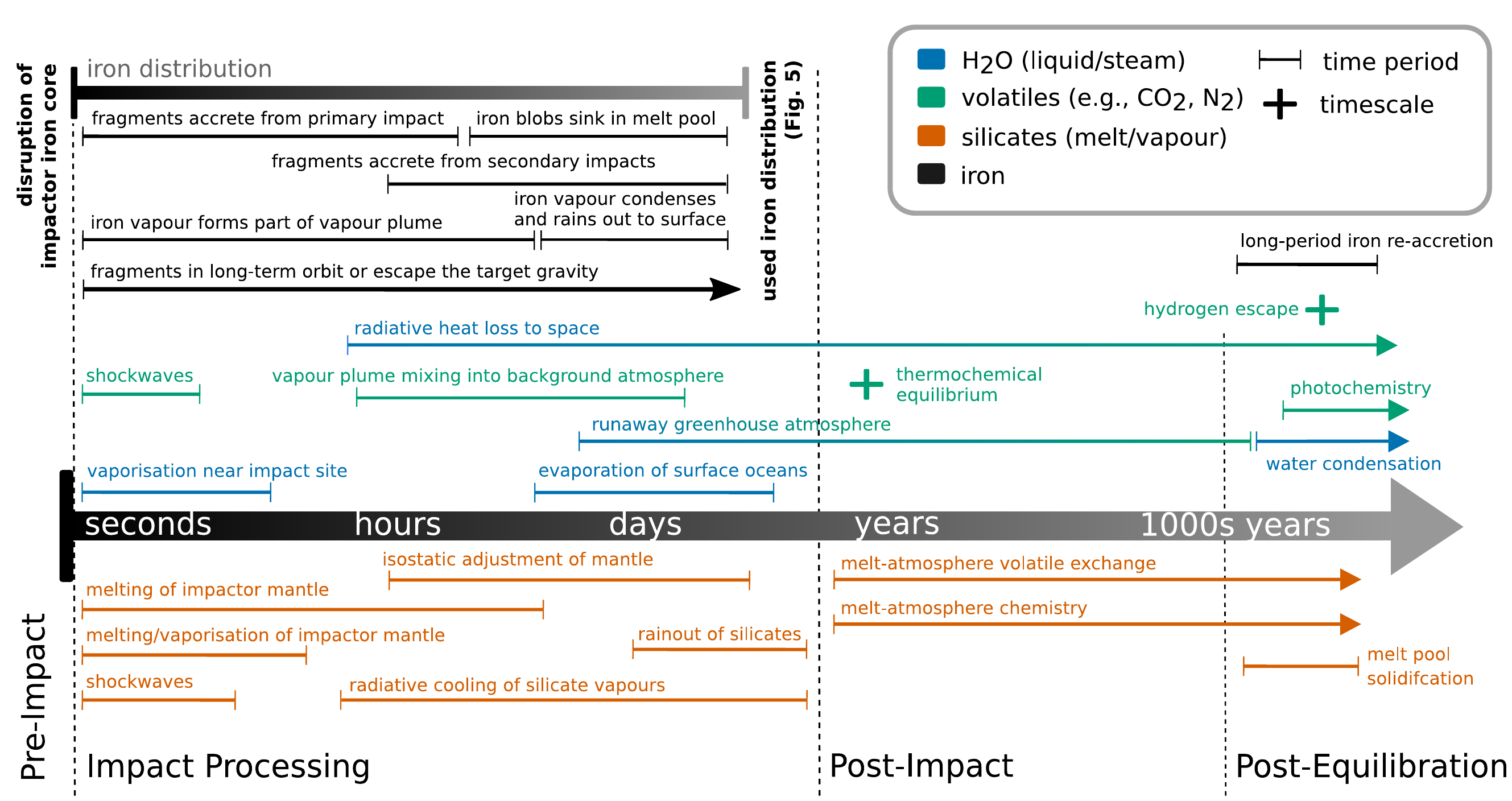}
    \caption{Timeline of important events and processes that the target and impactor experience from the moment of impact to long after it. The left-hand segment shows processes that we assume to have finished occurring by the time we begin equilibrium calculations involving melt-atmosphere interactions. The central segment shows the period of melt-atmosphere interactions, where the system is otherwise in a pseudo steady-state with few other processes occurring that substantially affect the state of the planet. The right-hand segment shows processes that we assume to occur after the melt and atmosphere have reached equilibrium, including solidification of the melt phase (throughout which melt-atmosphere interactions will also occur to maintain equilibrium, but a different equilibrium to that which we find in this study), hydrogen escape from the top of the atmosphere, significant cooling of the system to space, the resulting condensation of the steam atmosphere, and the ensuing photochemistry of \citet{zahnle2020creation}.}
    \label{fig:Timeline}
\end{sidewaysfigure}

Before calculations for melt-atmosphere interactions begin, we perform several calculations required to take the system from its state before the impact (Figure \ref{fig:Graphical Abstract}, \textit{left}) to its state in the aftermath (Figure \ref{fig:Cartoon}). We term these two states the ``pre-impact'' and ``post-impact'' states respectively. We term the processing of the impactor and target between these two states as ``impact processing'', which we estimate to take between under a year and several years (Figure \ref{fig:Timeline}) depending on the size of the impact. The state after subsequent equilibration between the atmosphere and melt phase will be termed the ``post-equilibration'' state.

This Section describes our impact processing calculations. We include a brief timeline of important processes which we consider to take place during this period. We then detail our treatments of these processes within the model, which is a simplified box model consisting of an atmosphere box and a melt phase box, with an interfacing boundary across which mass change occurs. We use a box model as the massive hot atmosphere should be compositionally well mixed, and for the large melt pool, which dominates most of our melt mass, the whole of the melt phase should cycle past the surface on timescales of weeks \citep{massol2016formation}. We group the impact processes into those dealing with production of the melt phase, those which affect the atmospheric composition, and those which determine the distribution of the impactor iron. Finally, we present the composition of the resulting post-impact system.

\input{sections/3.1_timeline}

\input{sections/3.2_melt_phase}

\input{sections/3.3_atmosphere}

\input{sections/3.4_metal}

\input{sections/3.5_post_impact}

%% file: sections/3.1_timeline.tex
\subsection{Events during impact processing}\label{sec:Processing Timeline}

We define our impact processing as beginning at the moment of contact between the impactor and the target atmosphere (Figure \ref{fig:Timeline}). This contact generates a shockwave which passes through the atmosphere, accelerating the gas and leading to atmospheric erosion. A second contact quickly follows ($\sim$ seconds later for an impactor travelling at $v_i \sim 20\,\mathrm{km~s^{-1}}$) between the impactor and the solid body of the target, producing a shock that travels through the two solid bodies with peak shock pressures $\gtrsim 100\,\mathrm{GPa}$.

In the minutes following contact, significant deformation of the target mantle occurs, with a large excavated region (scale of hundreds or thousands of km) forming in the wake of the impactor. The silicate materials of the target mantle experience substantial melting and vaporisation during the deformation and the rebound (i.e., adiabatic processing), much greater than that experienced directly from the shock. The impactor mantle and core experience substantial disruption; the mantle materials are almost entirely vaporised or melted, while the iron core experiences fragmentation, melting, and vaporisation (e.g., \citealt{genda2017terrestrial, citron2022large}). 

After tens of minutes, the initially chaotic structure of the system in the vicinity of and downstream from the impact site (i.e., along the impactor's initial velocity vector) settles out into a more defined structure. The melt pool starts to coalesce as the mantle rebounds, with iron either being suspended in the melt or sinking through it depending on a multitude of factors (e.g., size of iron fragments; see Section \ref{sec:Iron}). Above this melt, a mixed phase vapour plume forms. The plume consists predominantly of supercritical and vaporised silicates by mass, but also contains vaporised iron \citep{kraus2015impact}, suspended iron droplets, and volatiles stemming from the pre-impact atmosphere and those previously trapped in the silicate mantles of both the impactor and target. This mixed phase reaches thousands of degrees Kelvin \citep{svetsov2005numerical}. The surface water oceans in the vicinity of and downstream of the impact site are also vaporised at this stage. Away from the impact site, however, the target appears much as did it before the impact, with liquid water oceans and a \ce{CO2} dominated relatively cool atmosphere, although some impact-induced melting of the mantle does occur due to propagation of the impact shock.

On the scale of hours to days after impact, the vapour plume mixes into the background pre-impact atmosphere \citep{svetsov2005numerical}. The plume is meanwhile cooling, radiating heat upwards to space and downwards to the surface of the planet in approximately equal proportions at temperatures $\sim2000\,\mathrm{K}$ \citep{sleep1989annihilation}. The still-present surface water oceans efficiently absorb this downwards infrared radiation, and consequently evaporate (see Section \ref{sec:Vaporisation}) in the months to years following the impact, leading to the creation of a steam greenhouse atmosphere. Over this same time period, the silicates in the atmospheres cool sufficiently that they condense, forming droplets that rain down over the surface of the planet \citep{svetsov2005numerical}. Iron vapour in the atmosphere will similarly condense and rain down to the surface (Section \ref{sec:Iron}), and will likely start to do so before the silicates due to a higher condensation temperature. The silicate rain contributes to the mass of global melt, alongside melting induced by the surface temperature being above the silicates' solidus (e.g., Figure \ref{fig:Cartoon}). However, the mass contributions from both of these sources are small in comparison to the mass of melt produced directly in the impact (i.e., shock heating and adiabatic melting; see Section \ref{sec:Generating Melt}). Over the same time period, the target mantle undergoes isostatic adjustment, leading to the melt phase spreading slightly over the planet surface \citep{tonks1993magma}.

At the end of our impact processing calculations, therefore, we find a planet with a mostly solid mantle but with a large vigorously convecting melt pool of varying depth near the impact site plus a small layer of surface melt elsewhere (e.g., Figure \ref{fig:Cartoon}). The atmosphere is steam-dominated, with minor components from the pre-impact atmosphere and chemistry that has occurred (Section \ref{sec:Post-Impact}). Both the melt and atmosphere have been reduced chemically by interactions with the impactor iron they accreted during impact. The atmosphere is still hot, with temperatures at the melt-atmosphere interface (i.e., the planet ``surface'') being above the mantle solidus, and will take thousands of years to cool to its pre-impact temperature. The atmospheric temperature profile will be similar to profiles calculated by \citet{kasting1988runaway}, with lower atmosphere dry adiabats and temperatures decreasing to $< 500\,\mathrm{K}$ in the upper atmosphere. The phase boundary between the melt and the atmosphere is the so-called ``fuzzy layer'' \citep{kite2019superabundance}; it is not a distinct boundary but rather a transition layer where the phase and composition are ambiguous. In the model, however, we treat the boundary as distinct.

%% file: sections/3.2_melt_phase.tex
\subsection{Generating the melt phase}\label{sec:Generating Melt}
\subsubsection{Melt masses}\label{sec:Melt Masses}
We calculate the mass of the impact-generated melt phase using the GADGET2 \citep{springel2005cosmological}\footnote{Main code available at \url{https://wwwmpa.mpa-garching.mpg.de/gadget/}. Code adapted for use in planetary collisions, which was used in this study, is available in the supplement of \citet{cuk2012making}.} smoothed-particle hydrodynamics (SPH) code. We use the data presented in \citet{citron2022large}, where 48 simulations are carried out for varying impactor masses (0.0012, 0.0030, 0.0060, 0.0120 $M_\mathrm{Earth}$), impact velocities (1.1, 1.5, 2.0 $v_\mathrm{esc}$), and impact angles (0$^\circ$, 30$^\circ$, 45$^\circ$, 60$^\circ$). Impactors have a 1:2 mass ratio of iron core to silicate mantle. The target is Earth-like, with a $32.0\,\mathrm{wt}\%$ iron core and a $68.0\,\mathrm{wt}\%$ silicate mantle. The crusts of both bodies are neglected due to their small size relative to the particle resolution. All simulations were run for 24 hours of model time. Appendix \ref{sec:Appendix GADGET} contains further details of the simulation setup and the breakdown of how melt and vapour masses are determined from the mixed phase system, as well as the full data set.
\begin{figure}
    \centering
    \includegraphics[width=0.98\textwidth]{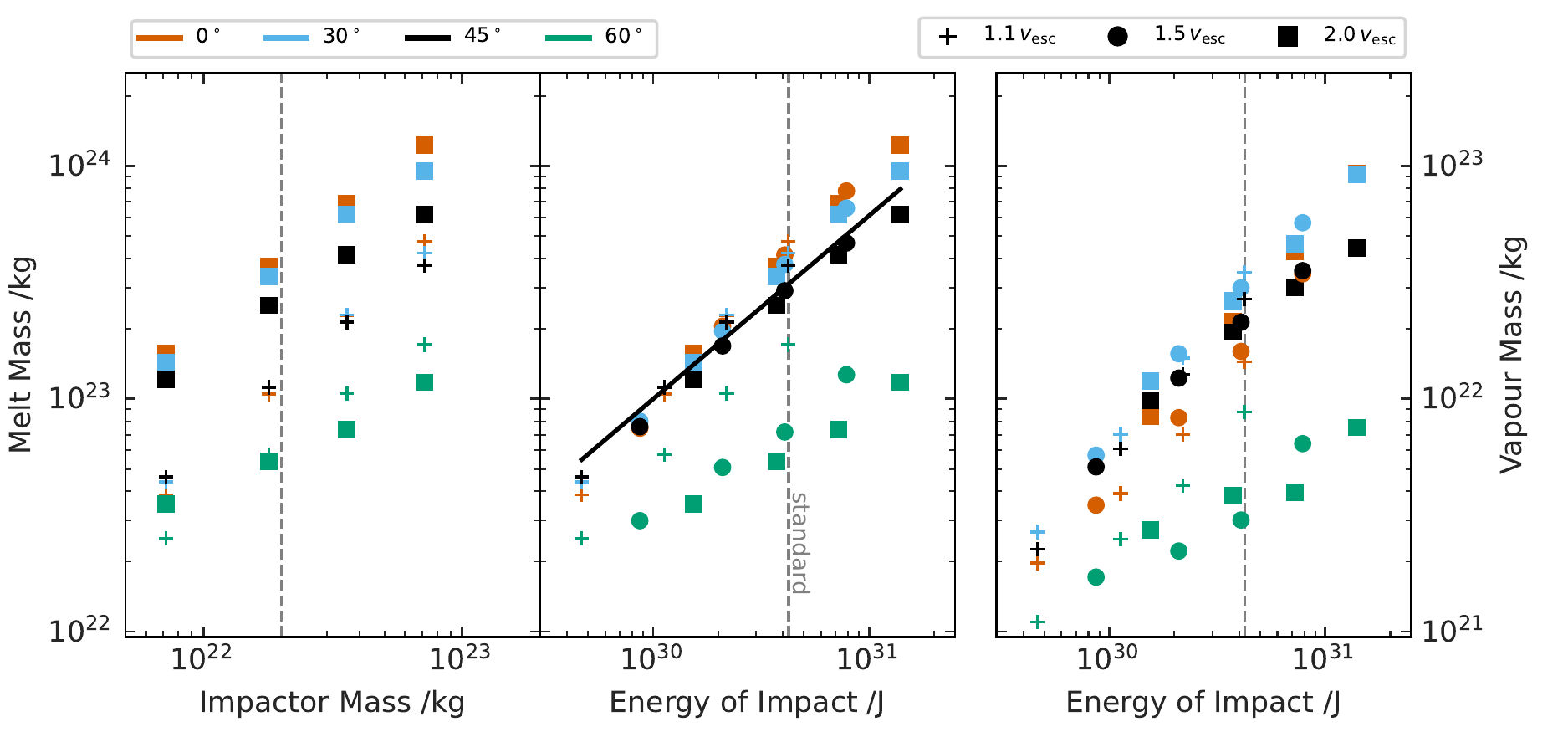}
    \caption{GADGET2 impact-generated melt and vapour masses as a function of impactor mass and impact energy. We show impact energy, despite using specific energy within our models, as this better demonstrates the increasing melt and vapour mass as the impacts become more head-on (i.e., lower $\theta_i$). The vertical dashed line indicates an impact under our standard values (Table \ref{tab:Standard Parameters}). The solid black line in the central panel is an example of how use linear fits of the data in log space to interpolate melt and vapour masses from the simulations. The total mass that we classify as our melt phase moving into melt-atmosphere interactions consists of the summed mass of melt and vapours (see text).}
    \label{fig:GADGET Melt Masses}
\end{figure}

We use our impact simulations to calculate the impact-generated melt mass as a function of the modified specific energy of impact \citep{stewart2015how}, 
\begin{equation}
    Q_S = Q_R \left( 1 + \frac{M_i}{M_t} \right) \left( 1 - b \right)~~,
    \label{eqn:Specific Energy}
\end{equation}
where $b$ is the impact parameter ($\sin\theta_i$ for impact angle $\theta_i$), and
\begin{equation}
    Q_R = \frac{1}{M_i + M_t} \left( \frac{1}{2}\mu v_i^2 \right)~~,
\end{equation}
\begin{equation}
    \mu = \frac{M_i M_t}{M_i + M_t}~~,
    \label{eqn:Reduced Mass}
\end{equation}
where $\mu$ is the reduced mass. We use simple linear interpolations of the simulation results in log space to determine melt and vapour masses as a function of $Q_S$ (Figure \ref{fig:GADGET Melt Masses}). The total mass of silicates that we use in our calculations of melt-atmosphere interactions consists of the total mass of melt and vapour as calculated by these interpolations, as the silicate vapours are expected to rain out to the surface during the period of impact processing (Figure \ref{fig:Timeline}). There would likely be a further contribution to the melt mass due to the melting of the planet surface under the extreme greenhouse atmosphere. We cannot accurately quantify this contribution, although we estimate that it will be small due to the large surface coverage of our calculated melt \citep{citron2022large}. As such, we consider our calculated melt masses as being the minimum melt mass produced in our impacts. Any additional melt mass will act to strengthen the effects of the physics we demonstrate in this study, especially as melt derived from the planet surface/crust would be closer in composition to basalt, which we show to have a greater oxidising effect than peridotite melts (Section \ref{sec:Results}).

In this study, as per our standard values (Table \ref{tab:Standard Parameters}), we focus on the melt produced in our $45^\circ$ impacts. These impacts produce less melt than the $0^\circ$ (head-on) and $30^\circ$ impacts for the same impactor mass and velocity, but much more melt than the $60^\circ$ impacts (Figure \ref{fig:GADGET Melt Masses}).

\subsubsection{Melt phase composition}\label{sec:Melt Composition}
The composition of the impact-generated melt is dependent on the completeness of melting \citep{kushiro1963origin}. Regions with high-fraction melting (melt fraction $\gtrsim 40\,\%$) will form melts with more peridotite-like composition, while regions with low-fraction melting (melt fraction $\lesssim 40\,\%$) will form melts with more basalt-like composition. Each of these compositions of melt phase will have a different $f$O$_2$ (Table \ref{tab:Standard Parameters}), and thus each have different redox buffering potential for the atmosphere.

Our simulations indicate that less than $\sim20\,\%$  of the melt generated during impact is derived from mantle experiencing low-fraction melting, while the remainder is from mantle experiencing high-fraction melting. The completeness of melting generally decreases with distance from the impact point as expected (e.g., \citealt{tonks1993magma}). We thus expect our impact-generated melt pool to be composed mostly of peridotitic melts with a large basaltic component. This basaltic component will be further contributed to by the melting of any basaltic crust, which is unresolved in our simulations. 

We present two end member cases in this study. In each case, we assume that the full impact-generated melt mass is composed of either a basaltic or peridotitic melt, and then calculate the equilibrium state of the interacting melt-atmosphere system. These single-composition melt phases are treated as well-mixed homogeneous bodies of melt. We discuss the likely contribution from each end member case to our impact scenarios in Section \ref{sec:Discussion}, considering the mobility and geometry of these two melt phase compositions, and the effect that this has on their ability to interact with the atmosphere.

%% file: sections/3.3_atmosphere.tex
\subsection{Atmospheric processing during impact}\label{sec:Atmos Processing}
\subsubsection{Atmospheric erosion} \label{sec:Erosion}
Each impact erodes the pre-impact atmosphere \citep{shuvalov2009atmospheric, schlichting2015atmospheric}. To calculate the mass lost, we follow the parametrisation of \citet{kegerreis2020atmospheric_b}, who used SPH simulations to derive scaling laws. The mass fraction of the atmosphere removed, $X$, is given by
\begin{equation}
    X \approx 0.64 \left[
    \left(\frac{v_i}{v_\text{esc}}\right)^2
    \left(\frac{M_i}{M_i + M_t}\right)^{1/2}
    \left(\frac{\rho_i}{\rho_t}\right)^{1/2}
    f(\theta_i, R_i, R_t) \right]^{0.65} \hspace{2mm},
    \label{eqn:Kegerreis Mass Loss}
\end{equation}
where $v_i$ and $v_\text{esc}$ are the impact and escape velocities, $\rho_i$ and $\rho_t$ are the impactor and target bulk densities (not considering the atmosphere), and $f(\theta, R_i, R_t)$ is the fraction of interacting mass between the impactor and target of radii $R_i$ and $R_t$. Atmospheric mass loss from impacts does not lead to fractionation of gas species \citep{schlichting2018atmosphere}. In our box model, therefore, mass ejection simply removes equal mass fraction, $X$, of all species in our pre-impact atmosphere box.

\subsubsection{Volatiles and vapours} \label{sec:Volatiles}
We assume that the silicate and iron vapours in the impact-generated vapour plume condense and rain out of the atmosphere during impact processing (Section \ref{sec:Processing Timeline}). Our treatment of the iron is detailed in Section \ref{sec:Iron}. For the silicates, chemistry could occur with the volatiles in the ambient atmosphere before the vapours condense. However, the silicate phases are likely oxidising in nature \citep{kuwahara2015molecular}, and given the inherent oxidising power of the impact-vaporised steam atmosphere, we anticipate the additional oxidising power of the silicate phases to be relatively small. We thus do not account for the additional oxidising power of the silicate vapours in the composition of the post-impact atmosphere. 

The silicate mantles of both the impactor and target will contain volatile species. Upon vaporisation of the silicates, these volatiles will be released into the atmosphere. We treat this release of volatiles within our model by assuming that the water content of the impactor mantle is fully released into the atmosphere, constituting to a few bars or $\sim0.01\,\mathrm{EO}$. We recognise that not all of the impactor mantle will be vaporised and have its volatiles released in this way, in addition to the fact that the target mantle will also experience vaporisation. However, the mass of volatiles released is so small in comparison to the mass of the atmosphere that the error is likely insubstantial.

\subsubsection{Ocean vaporisation} \label{sec:Vaporisation}
There is sufficient energy in a $\sim 10^{21}\,\mathrm{kg}$ impactor, travelling at $1.5\,v_{esc}$ and impacting at $45^\circ$ to vaporise an Earth Ocean's worth of water solely through the internal energy increase of the atmosphere, even after accounting for approximately half the energy radiating to space \citep{citron2022large}. If the internal energy increase of the target's surface layer is also accounted for, and this energy is efficiently transported to the atmosphere, a $\sim 10^{19}\,\mathrm{kg}$ impactor would suffice. We thus assume that all our impactors ($2\times10^{21}-2\times10^{22}\,\mathrm{kg}$) are capable of vaporising our pre-impact 1.85 EO surface water budget. The vaporisation of the target surface oceans injects hundreds of bars of water vapour into our atmospheres. 

\subsubsection{Iron interaction and atmospheric chemistry} \label{sec:FastChem} 
The impactor iron made available to the atmosphere (see Section \ref{sec:Iron}) chemically reduces the atmosphere. We treat this by stoichiometrically reducing \ce{H2O} into \ce{H2}, ignoring the reduction of other oxidised species in the atmosphere (e.g., \ce{CO2}). We do this as we place the atmosphere in thermochemical equilibrium as the final step of our impact processing of the atmosphere, through which the iron's reducing power affects all atmospheric species regardless, as if they were included in the original stoichiometric reduction. 

The addition of volatiles into the pre-impact atmosphere throughout the stages impact processing leads to a gas mixture that is out of equilibrium. Our final processing step is thus to calculate the composition of the atmosphere under thermochemical equilibrium, for which we use the \texttt{FastChem}\footnote{available at \url{https://github.com/exoclime/FastChem}} package \citep{stock2018fastchem} . We further verified the results using our own chemical kinetics network solver, based on the model presented in \citet{hobbs2019chemical}. We use a box temperature of $1900\,\mathrm{K}$ for our calculations as any produced silicate vapours will have condensed and rained out at this temperature \citep{svetsov2005numerical}, and silicates will still be molten at the planet surface \citep{sossi2020redox}, approximately corresponding to the end of our impact processing stage (Figure \ref{fig:Timeline}).

%% file: sections/3.4_metal.tex
\subsection{Distribution of the impactor iron}\label{sec:Iron}
The iron cores of large impactors stretch and fragment during impact. Molten iron fragments $\sim10\,\mathrm{m}$ in diameter form \citep{genda2017terrestrial}, which then further break up into $\sim1\,\mathrm{m}$ blobs if accreted during the initial collision \citep{kendall2016differentiated}, or can form a $\sim1\,\mathrm{mm}$ hail if accreted during secondary impacts of initially scattered iron fragments \citep{genda2017terrestrial}. The iron can also experience sufficiently high shock heating as to be vaporised during the impact \citep{kraus2015impact}, forming part of the impact-generated vapour plume (Section \ref{sec:Processing Timeline}). Alternatively, some iron fragments have sufficient kinetic energy to escape the system \citep{genda2017ejection, marchi2018heterogeneous, citron2022large}.

We use our GADGET2 SPH simulations to calculate the distribution of the impactor iron within the system. We divide the iron into 3 reservoirs based on its location 24 hours of model time into the GADGET simulations (see Appendix \ref{sec:Appendix GADGET} for details of how particles are classified within the SPH simulations as belonging to which reservoir):
\begin{enumerate}
    \item{Atmosphere - iron that is accreted by the target atmosphere, consisting mainly of iron that is vaporised during the impact or that forms part of the impact ejecta. We make this iron available to reduce \ce{H2O} in the atmosphere to \ce{H2} (Section \ref{sec:Iron});}   
    
    \item{Interior - iron that is accreted by the target interior. Portions of the iron in this reservoir (generally smaller fragments) will emulsify into small droplets and be available to chemically reduce the melt phase. Other portions (generally larger fragments) will sink to bottom of the melt pool, and although they experience turbulent mixing with the melt while they sink (e.g., \citealt{rubie2003mechanisms, dahl2010turbulent, deguen2014turbulent, kendall2016differentiated}), the iron they contain is less available to interact chemically with the melt phase, if at all. 
    
    Improved impact simulations are required to better capture these processes, and hence the characterisation of iron accretion by the target interior. In particular, models with appropriate elastic-plastic interior rheologies are required. Without such models, we present two extreme cases: in Model 3A, all of the iron accreted by the target interior is available to reduce the melt phase, either immediately as the iron accretes, or later during melt-atmosphere interactions through iron droplets remaining in the vigorously convecting melt; in Model 3B, none of the interior iron is available to reduce the melt phase. The most accurate distribution will likely be somewhere in between these two extremes (Section \ref{sec:Discussion Iron}); and}
    
    \item{Not Accreted - iron that dynamically escapes the system or that ends up in orbit around the target in a disk. We assume that this iron does not interact with the target within the lifetime of the impact-generated melt phase. Instead, it is re-accreted on Myr timescales, the consequences of which are beyond the scope of this work.}
\end{enumerate}
\begin{figure*}
    \centering
    \includegraphics[width=0.94\textwidth]{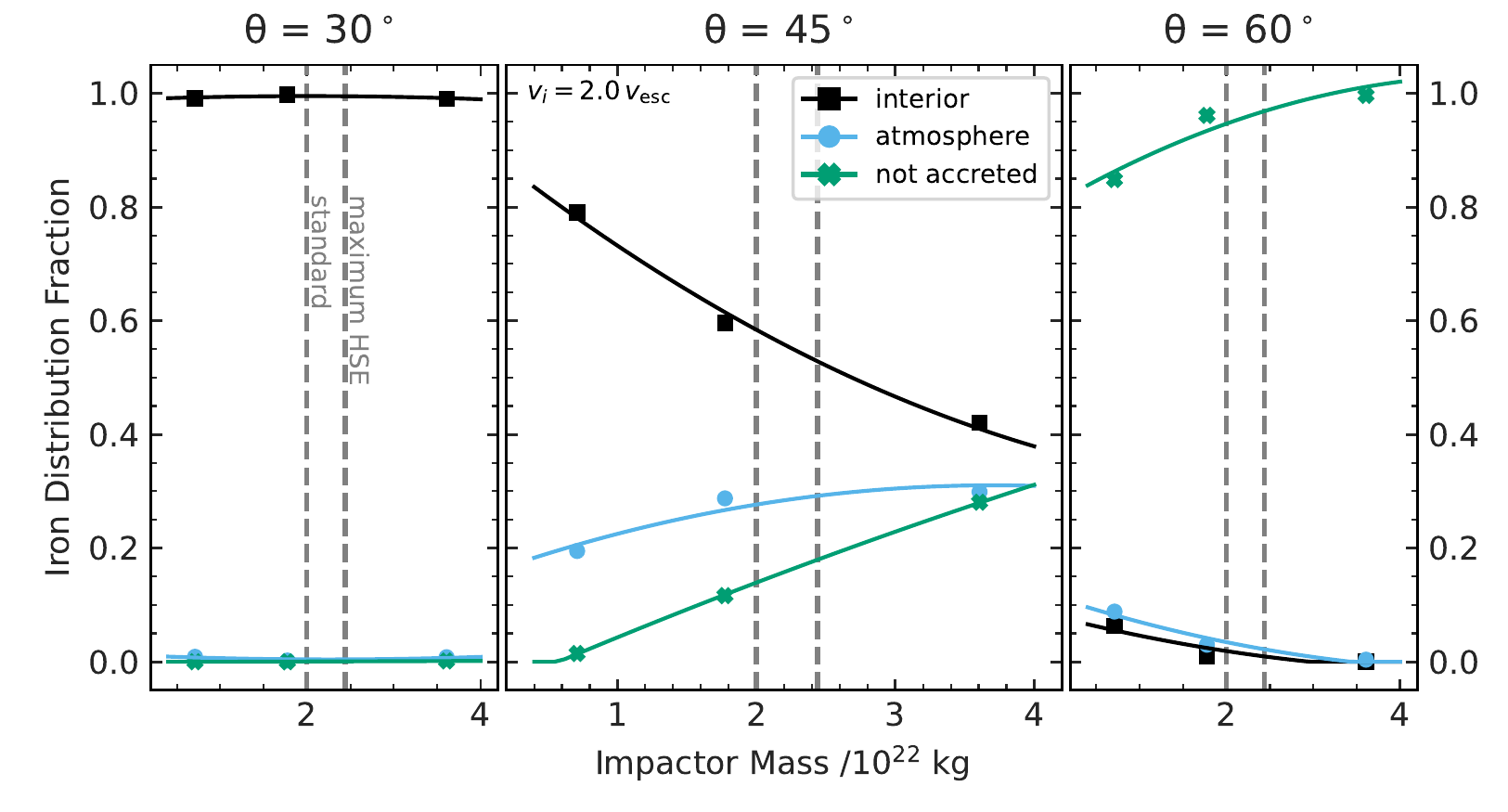}
    \caption{Distribution of impactor iron inventory between different target reservoirs as a function of impactor mass for $v_i=2v_\text{esc}$. The solid lines are quadratic fits in log-linear space that we use to interpolate iron distributions from our simulations.}
    \label{fig:Iron Distribution}
\end{figure*}

The iron distribution between reservoirs changes greatly with impactor mass, and also with impact angle (Figure \ref{fig:Iron Distribution}). For our standard impact velocity ($v_i = 2~v_{esc}$) and impact angle ($\theta_i = 45^\circ$), we observe that the fraction of iron deposited into the target interior decreases rapidly with increasing $M_i$, while the fraction of iron that escapes the system increases. The fraction of iron available to the atmosphere remains in the range $20-30\,\mathrm{\%}$. For impactor masses not explicitly modelled, we interpolate between simulated values (Figure \ref{fig:Iron Distribution}). For $\theta_i = 30^\circ$, almost all of the impactor's iron inventory is delivered to the target interior for all impactor masses, and an even greater fraction is delivered there for head-on ($\theta_i = 0^\circ$) impacts. This is expected behaviour based on 2D simulations of head-on collisions (e.g., \citealt{kendall2016differentiated}). For $\theta_i = 60^\circ$, a large proportion of the iron remains unaccreted by the target, with this proportion increasing with impactor mass (Figure \ref{fig:Iron Distribution}).

In our box model, we assume that all iron made available to the atmosphere during accretion reacts completely with the \ce{H2O} in the atmosphere. This treatment represents an upper limit on the iron that can react with the atmosphere, as some iron may condense and rain out to the surface before interaction (Section \ref{sec:Processing Timeline}). Calculating the interacting fraction is beyond the scope of our models, however, and we have no known way of estimating it. Further, as we show in Section \ref{sec:Results}, as long as the reducing power of the iron is accreted by either the melt phase or the atmosphere, melt-atmosphere interactions mean that this reducing power can influence the equilibrium composition of the atmosphere. Iron raining out from the atmosphere will likely accrete to the melt phase in small enough fragments (e.g., droplets) that it will readily react with the melt and thus will be able contribute to the redox state of the equilibrium atmosphere. Our inability to treat iron rainout should hence have relatively little effect on the results we present in this study for models which include melt-atmosphere interactions. The iron reacts with atmospheric \ce{H2O} via
\begin{equation}
    \ce{Fe + H2O -> H2 + FeO}~~.
    \label{eqn:Reduction}
\end{equation}
The \ce{FeO} produced in this reaction condenses and rains out to the surface before the end of the impact processing period. We thus add the moles of \ce{FeO} produced directly to the impact-generated melt phase of Section \ref{sec:Generating Melt}. The iron that we make available to reduce our melt phase rapidly reacts with the melt \citep{takada1986determination} via
\begin{equation}
    \ce{Fe + Fe2O3 -> 3FeO}~~,
    \label{eqn:Re-equilibration}
\end{equation}
with the reaction terminating when there is either no metallic iron left or the melt phase becomes metal saturated (see Section \ref{sec:Oxygen Fugacity}).

%% file: sections/3.5_post_impact.tex
\subsection{The post-impact system} \label{sec:Post-Impact}
\begin{figure*}
    \centering
    \includegraphics[width=0.99\textwidth]{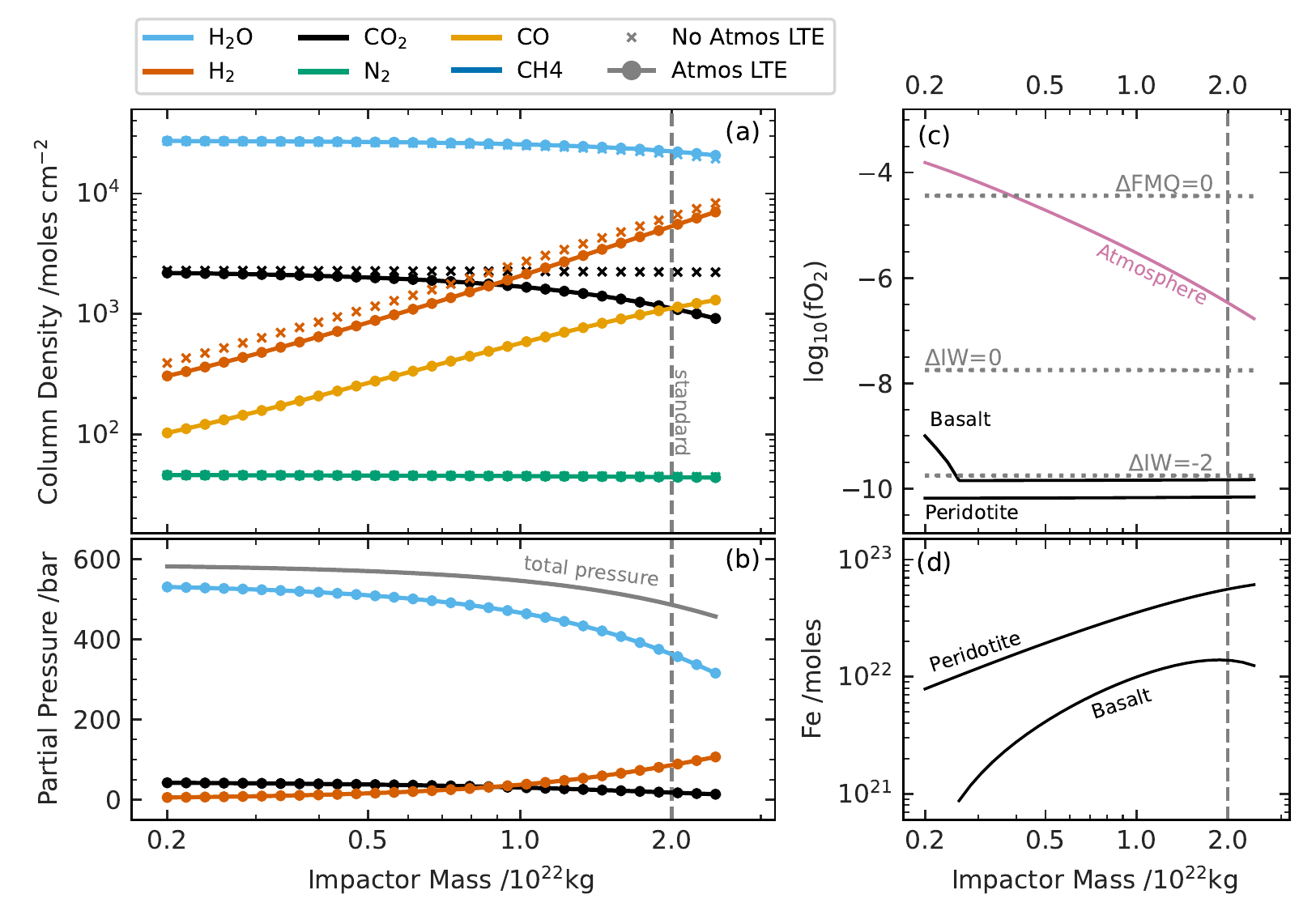}
    \caption{The state of the system at the post-impact stage as a function of impactor mass in Model 3A. All model parameters except for the varying impactor mass are the standard values detailed in Table \ref{tab:Standard Parameters}. (a) Composition of the atmosphere with (circles) and without (crosses) the final processing step of local thermochemical equilibrium (LTE). (b) Partial pressures of the major 3 species in the atmosphere, and total atmospheric pressure. (c) Oxygen fugacities of the atmosphere, basaltic melt and peridotitic melt. The FMQ and iron w\"ustite (IW) buffers are shown for reference (dotted lines). (d) Metallic iron remaining in the melt pool after metal saturation has been reached for each melt phase composition. Appendix \ref{sec:Appendix post-impact} contains a similar Figure for Model 3B.}
    \label{fig:Init Atmos}
\end{figure*}

The composition of the post-impact atmosphere changes greatly with impactor mass (Figure \ref{fig:Init Atmos}a,b). Although steam dominates the atmosphere in almost all cases, as $M_i$ increases, there is proportionally more iron due to the impactors' fixed iron mass fraction and similar fractions of the iron being accreted by the atmosphere. The post-impact atmosphere thus contains substantially more \ce{H2} as $M_i$ increases. The remnant \ce{CO2} from the pre-impact atmosphere is still a major species in the atmosphere for lower impactor masses, at $\lesssim10\,\%$. However, as the atmosphere becomes more reducing with increasing $M_i$, \ce{H2} overtakes \ce{CO2} as the atmosphere's secondary species. This is due to both the increasing \ce{H2} abundance and the rise of \ce{CO} as a carbon-bearing species through reducing atmospheric chemistry. In the most reducing atmospheres, we observe further reduction of carbon to form small quantities of \ce{CH4}. \ce{N2} does not change significantly with impactor mass, although we do observe small quantities of \ce{NH3} in the most reducing atmospheres (not visible on Figure \ref{fig:Init Atmos}a). The total atmospheric pressure decreases as $M_i$ increases, as the oxidation of Fe to FeO and subsequent rainout of the FeO to the surface represents a mass loss from the atmosphere. The atmospheric $f$O$_2$ becomes monotonically more reduced with increasing $M_i$ (Figure \ref{fig:Init Atmos}c), in line with the compositional changes seen.

Under Model 3A, where the iron accreted by the interior reduces the melt phase, all impactors deliver sufficient iron to the melt to cause metal saturation if the melt is purely peridotitic in composition. We thus observe a flat curve of $f$O$_2$ against $M_i$ (Figure \ref{fig:Init Atmos}c). For a wholly basaltic impact-generated melt phase, only the smallest impactors are not able to saturate the melt in metallic iron. For metal saturated melts, any accreted iron not used up in reducing the melt phase goes into forming a metal phase within the magma. The increasing total iron inventory with increasing $M_i$ competes with the decreasing efficiency of accretion by the target interior, as well as the logarithmically increasing melt mass, to produce the curves seen in Figure \ref{fig:Init Atmos}d. Generally, however, the amount of iron in the metal phase increases with $M_i$. The exception to this is again for the smallest impactors and a basaltic melt phase composition, where we observe no metallic iron at the post-impact stage (Figure \ref{fig:Init Atmos}d). For a given impactor mass, there is a greater reservoir of metallic iron remaining in the peridotitic melts than the basaltic melts. This is due to the peridotitic melts being more reducing at the point of melting, and thus less iron being consumed in the peridotite reaching metal saturation than for the basalt. 

In the case of Model 3B (Appendix \ref{sec:Appendix post-impact}), impactor iron is not available to reduce the melt phase, meaning that the melt phase redox state is equal to that at which it formed (2.3 log units below the FMQ buffer for the peridotitic melt and $\Delta\mathrm{FMQ}=0$ for the basaltic; Table \ref{tab:Standard Parameters}), and there is no chemically available metallic phase the magma (i.e., zero moles of Fe in Figure \ref{fig:Init Atmos}d equivalent). The atmospheres of Model 3A \& 3B are identical at the post-impact stage.

%% file: sections/4_equilibration.tex
\section{Melt-atmosphere equilibration}\label{sec:Equilibration}
The post-impact atmosphere and melt phase interact until equilibrium. Under the box model set out in Section \ref{sec:Processing Timeline}, chemical reactions occur between species in the gas phase (\ce{H2O} and \ce{H2} in the atmosphere box) and species in the melt and metal phases (\ce{Fe2O3}, \ce{FeO}, and Fe in the magma box). Partitioning of \ce{H2O} between the gas phase and the melt phase also occurs. We define equilibrium as having been achieved when the oxygen fugacities of the melt and atmosphere are equal, and simultaneously the partial pressures of \ce{H2O} in both reservoirs are also equal.

In our calculations, we enforce melt-atmosphere equilibrium in a step-wise time-independent manner. The first step towards equilibrium is a redox step, which takes the system to $f$O$_2$ equilibrium. This is followed by an \ce{H2O} partitioning step which takes the system to \ce{H2O} equilibrium. These two steps repeat until both properties converge. This Section details the calculations involved in performing such steps, which are identical in all model versions that include melt-atmosphere interactions. 

\subsection{Oxygen fugacity}\label{sec:Oxygen Fugacity}
The parametrisation which we use to describe the melt phase $f$O$_2$ is dependent on whether or not the melt is metal saturated. We take metal saturation as occurring at 2 log units below the iron w\"ustite (IW) buffer \citep{righter2012redox}. When a peridotitic melt is metal unsaturated, we calculate its $f$O$_2$ following the prescription of \citet{sossi2020redox}. They measured the dependence of the ferrous-ferric equilibrium on $f$O$_2$ in silicate melts of bulk silicate Earth (BSE) composition under conditions of $1\,\mathrm{bar}$ and $2173\,\mathrm{K}$. The equilibrium can be written as
\begin{equation}
    \ce{Fe^{3+}O_{1.5} <=> Fe^{2+}O + 1/4 O_2}~~,
    \label{eqn:FMQ Equilibrium}
\end{equation}
or alternatively written as a heterogeneous melt-gas phase reaction,
\begin{equation}
    \ce{Fe_2O_3$^\text{melt}$ + H_2$^\text{gas}$ <=> 2FeO$^\text{melt}$ + H_2O$^\text{gas}$}~~.
    \label{eqn:FMQ Reaction}
\end{equation}
\citet{sossi2020redox} developed the formulation
\begin{equation}
    \log_{10} \bigg(\frac{X_\mathrm{Fe^{3+}O_{1.5}}}{X_\mathrm{Fe^{2+}O}} \bigg) = (0.252 \pm 0.015) \Delta\mathrm{IW} - (1.530 \pm 0.053)~~,
    \label{eqn:Sossi}
\end{equation}
where $X_\mathrm{Fe^{3+}O_{1.5}}$ and $X_\mathrm{Fe^{2+}O}$ are molar fractions of ferric and ferrous iron in the melt phase respectively, and $\Delta\mathrm{IW}$ is the melt phase $f$O$_2$ relative to the IW buffer \citep{frost1991introduction}. 

When a basaltic melt is metal unsaturated, we calculate its $f$O$_2$ following the prescription of \citet{kress1991compressibility}, an empirical formulation that is based on measurements of the ferrous-ferric equilibrium by \citet{kennedy1948equilibrium}, \citet{fudali1965oxygen}, \citet{thornber1980effect}, \citet{sack1981ferric}, \citet{kilinc1983ferric}, \citet{kress1988stoichiometry}, and \citet{kress1991compressibility} between temperatures of $927-1357\,\mathrm{K}$:
\begin{equation}
    \ln \bigg(\frac{\mathrm{X}_{\mathrm{Fe}_{2}\mathrm{O}_{3}}}{\mathrm{X}_{\mathrm{FeO}}} \bigg) = a\ln\left(f\mathrm{O}_{2}\right)~+~\frac{b}{T}~+~c~+~\sum_{i} d_{i} X_{i} +~e\Bigg[1 - \frac{T_0}{T} - \ln\left(\frac{T}{T_0}\right)\Bigg] + f~\frac{P}{T}~+ g~\frac{(T-T_0)P}{T} + h~\frac{P^2}{T}~~,
\label{eqn:Kress and Carmichael}
\end{equation}
where $X_\text{\ce{Fe2O3}}$ and $X_\text{\ce{FeO}}$ are molar fractions, and prescription parameters $a$ to $h$ are given in Table \ref{tab:Kress and Carmichael}. Temperatures are given in units of K, pressures in units of Pa, and $f\mathrm{O}_2$ in units of bars.  
\begin{table}
    \centering
    \caption{Parameter values for the $f\!\mathrm{O}_2$ prescription of \citet{kress1991compressibility} in Equation \ref{eqn:Kress and Carmichael}. FeO$^*$ represents all oxidised states of iron in the melt.}
    \label{tab:Kress and Carmichael}
    \begin{tabular}{c|l|||c|l}
        \toprule
        a & ~0.196 & $d_\text{\ce{Al_2O_3}}$ & -2.243 \\
        b & ~1.1492 $\times 10^{4}\,\mathrm{K}$ & $d_\text{~FeO$^*$}$ & -1.828 \\
        c & -6.675 & $d_\text{CaO}$ & ~3.201 \\
        e & -3.36 & $d_\text{\ce{K_2O}}$ & ~6.215 \\
        f & -7.01 $\times 10^{-7}\,\mathrm{K~Pa^{-1}}$ & $d_\text{\ce{Na_2O}}$ & ~5.854 \\
        g & -1.54 $\times 10^{-10}\,\mathrm{Pa^{-1}}$ & \quad \\
        h & ~3.85 $\times 10^{-17}\,\mathrm{K~Pa^{-2}}$ & $T_0$ & ~$1673\,\mathrm{K}$ \\
        \bottomrule
    \end{tabular}
\end{table}

When the melt is metal saturated, we must also consider the equilibrium between ferrous and metallic iron, 
\begin{equation}
    \ce{2Fe^2+O <=> 2Fe^0 + O_2}~~,
    \label{eqn:IW Equilibrium}
\end{equation}
or alternatively written as the reaction,
\begin{equation}
    \ce{FeO$^\text{melt}$ + H_2$^\text{gas}$ <=> Fe^$\text{metal}$ + H_2O$^\text{gas}$}~~.
    \label{eqn:IW Reaction}
\end{equation}
At metal saturation, when there is little \ce{Fe2O3} present, we switch from tracking the melt phase $f$O$_2$ via Equations \ref{eqn:Sossi} \& \ref{eqn:Kress and Carmichael} to an alternative prescription. We use the formulation of \citet{frost1991introduction} to describe how the $f$O$_2$ of the IW buffer changes with pressure and temperature, and the formulation of \citet{righter2012redox} to describe how the melt phase $f$O$_2$ changes relative to the IW buffer with varying ratio of \ce{Fe} in the metal phase and \ce{FeO} in the silicate melt. Combining these two prescriptions, we have
\begin{equation}
        \log_{10}(f\mathrm{O}_2) = -\frac{2748}{T} + 6.702 + 0.055\left(\frac{P - 1}{T}\right) -2\log_{10}\left(\frac{X_\text{Fe}}{X_\text{FeO}}\right)
    \label{eqn:Frost}
\end{equation}
where $T$ is given in units of K, and $P$ is now given in units of bars. $X_\text{Fe}$ is the molar fraction of iron in the metal phase, which we take to be a fixed value of 0.98 (i.e., an almost purely iron metal phase).

The $f$O$_2$ prescriptions for both Equations \ref{eqn:FMQ Equilibrium} \& \ref{eqn:IW Equilibrium} are based on chemical equilibria which take place in the presence of hydrogen but do not explicitly include it. The chemical reactions (Equations \ref{eqn:FMQ Reaction} \& \ref{eqn:IW Reaction}) are what stoichiometrically drive the redox chemistry in the system. Thus, to complete the characterisation of our system, we require a final equilibrium expression, 
\begin{equation}
    \ce{H_2 + 1/2 O_2 <=> H_2O}~~,
    \label{eqn:Atmos Equilibrium}
\end{equation}
the equilibrium constant for which can be expressed as
\begin{equation}
    k_\text{eq}^\text{atm} = \frac{f\ce{H_2O}}{f\ce{H_2} (f\mathrm{O}_2)^{1/2}}~~, 
    \label{eqn:Atmos Equilibrium Constant}
\end{equation}
where $f$H$_2$ and $f$H$_2$O are the fugacities of hydrogen and water in the atmosphere. Assuming an ideal gas, $f$H$_2 \approx X_{\ce{H2}}$, and the atmospheric $f$O$_2$ can then be written as
\begin{equation}
    -1/2\log_{10}(f\mathrm{O}_2) = \log_{10}(k_\text{eq}^\text{atm}) + \log_{10}\left(\frac{X_{\ce{H2}}}{X_{\ce{H2O}}}\right)~~,
    \label{eqn:Atmos Fugacity}
\end{equation}
for atmospheric mixing ratios $X_i$. We take the equilibrium constant following the prescription of \citet{ohmoto1977devolatilization},
\begin{equation}
    \log_{10}(k_\text{eq}^\text{atm}) = \frac{12510}{T} - 0.979 \log_{10}(T) + 0.483 ~~.
    \label{eqn:Ohmoto Equilibrium Constant}
\end{equation}

\subsection{$f$O$_2$ equilibrium} \label{sec:fO2 Equilibrium}
At the beginning of each redox step, the atmosphere can either be in a more reducing or more oxidising state than the melt phase. 

\subsubsection{Reducing atmosphere case ($f$O$_2^\mathrm{atmos} < f$O$_2^\mathrm{melt}$)}
While $f$O$_2^\mathrm{melt} > \Delta\mathrm{IW-2}$, the melt phase is unsaturated in metal. We thus stoichiometrically reduce \ce{Fe2O3} to \ce{FeO}, which has the effect, on considering the interaction with the atmosphere, of consuming \ce{H2} and producing \ce{H2O}. The melt phase $f$O$_2$ in this region is calculated via Equation \ref{eqn:Sossi} for a peridotitic melt or Equation \ref{eqn:Kress and Carmichael} for a basaltic melt. If we do not reach $f$O$_2$ equilibrium through this stoichiometric reduction, we eventually reach a melt phase 2 log units below the IW buffer. At $\Delta\mathrm{IW}-2$, we switch to simultaneous reduction of \ce{Fe2O3} to \ce{FeO} and \ce{FeO} to \ce{Fe}. Detailed calculations for simultaneous reduction are derived in Appendix \ref{sec:Appendix Simultaneous}. During simultaneous reduction, the melt phase $f$O$_2$ is still calculated via either Equation \ref{eqn:Sossi} or \ref{eqn:Kress and Carmichael}, switching to calculation via Equation \ref{eqn:Frost} when the relevant prescription predicts equal $f$O$_2$ to Equation \ref{eqn:Frost}. Upon switching, we start the sole reduction of \ce{FeO} to \ce{Fe} until the melt obtains $f$O$_2$ equilibrium with the atmosphere.

\subsubsection{Oxidising atmosphere case ($f$O$_2^\mathrm{atmos} > f$O$_2^\mathrm{melt}$)}
Instead of using up reducing power in the atmosphere (available atmospheric \ce{H2}) we now consume oxidising power (available atmospheric \ce{H2O}) as we proceed towards melt-atmosphere equilibrium. In cases where the post-impact melt phase is metal saturated, and we are in the regime where Equation \ref{eqn:Frost} is applicable, we oxidise metallic Fe to FeO only until Equation \ref{eqn:Frost} predicts the same melt phase $f$O$_2$ as either Equation \ref{eqn:Sossi} (for a peridotitic melt) or Equation \ref{eqn:Kress and Carmichael} (for a basaltic melt), or until the melt-atmosphere system reaches $f$O$_2$ equilibrium. If not in equilibrium, we then carry out simultaneous oxidation of Fe to FeO and FeO to \ce{Fe2O3} (Appendix \ref{sec:Appendix Simultaneous}) until the Fe metal is depleted or melt-atmosphere equilibrium is reached. The melt phase $f$O$_2$ is calculated using either Equation \ref{eqn:Sossi} or \ref{eqn:Kress and Carmichael} in this regime. Once the metal is depleted, we oxidise FeO to \ce{Fe2O3} until melt-atmosphere $f$O$_2$ equilibrium. In practice, FeO is never depleted.

\subsection{\ce{H2O} partial pressure}\label{sec:H2O Equilibration}
In partitioning \ce{H2O} between the melt phase and the atmosphere, we adopt the formulation of \citet{carroll1994volatiles}, which gives the saturation vapour pressure of \ce{H2O} over a melt phase with a given volatile mass fraction, based on the Burnham \ce{H2O} solubility model \citep{burnham1974role},
\begin{equation}
    X_{\ce{H2O}} = 6.8\times10^{-8} (p\ce{H2O}_\text{atm})^{0.7}~~,
    \label{eqn:H2O Equilibrium Calc}
\end{equation}
where $p\ce{H2O}_\text{atm}$ (units of Pa) is the atmospheric partial pressure of \ce{H2O} expected at equilibrium with a melt phase with an \ce{H2O} mass fraction $X_{\ce{H2O}}$. This formulation is based off of measurements of \ce{H2O} saturation in plagioclase feldspars (see Figure 1 of \citealt{burnham1974role}) for temperatures between $973-1173\,\mathrm{K}$ and pressures up to $10\,\mathrm{kilobars}$. We use Equation \ref{eqn:H2O Equilibrium Calc} to transfer \ce{H2O} between the atmosphere and the melt phase, either through dissolution or outgassing, until the \ce{H2O} partial pressure is equal across the boundary layer. Our simplified box model assumes that water can rapidly dissolve into or outgas from the melt phase, and that the convective timescales within the melt are rapid enough that the melt remains well-mixed in its volatile content. 

We do not account for the partitioning of other volatiles between the atmosphere and the melt phase. The solubilities of other species (e.g., \ce{CO2}, \ce{H2}) are orders of magnitude lower than that of \ce{H2O} at the atmospheric pressures in our study\footnote{See Figure 2 of \citealt{lichtenberg2021vertically} for a comparison of solubility data for \ce{H2O}: \citet{silver1990influence, holtz1995h2o, moore1998hydrous, yamashita1999experimental, gardner1999experimental, liu2005solubility}; \ce{H2}: \citet{gaillard2003rate, hirschmann2012magma}; \ce{CO2}: \citet{mysen1975solubility, stolper1988experimental, pan1991pressure, blank1993solubilities, dixon1995experimental}; \ce{CH4}: \citet{ardia2013solubility, keppler2019graphite}; \ce{N2}: \citet{libourel2003nitrogen, li2013nitrogen, dalou2017nitrogen, mosenfelder2019nitrogen}; \ce{CO}: \citet{yoshioka2019carbon}.}, and as such, including their partitioning would have little to no effect on the atmospheric composition during melt-atmosphere interactions.

%% file: sections/5_results.tex
\section{Results}\label{sec:Results}
\input{sections/5.1_standard}

\input{sections/5.2_model_comparisons}

\input{sections/5.3_remixing}

%% file: sections/5.1_standard.tex
\subsection{Standard values system}\label{sec:Standard Result}
\begin{figure*}
    \centering
    \includegraphics[width=0.98\textwidth]{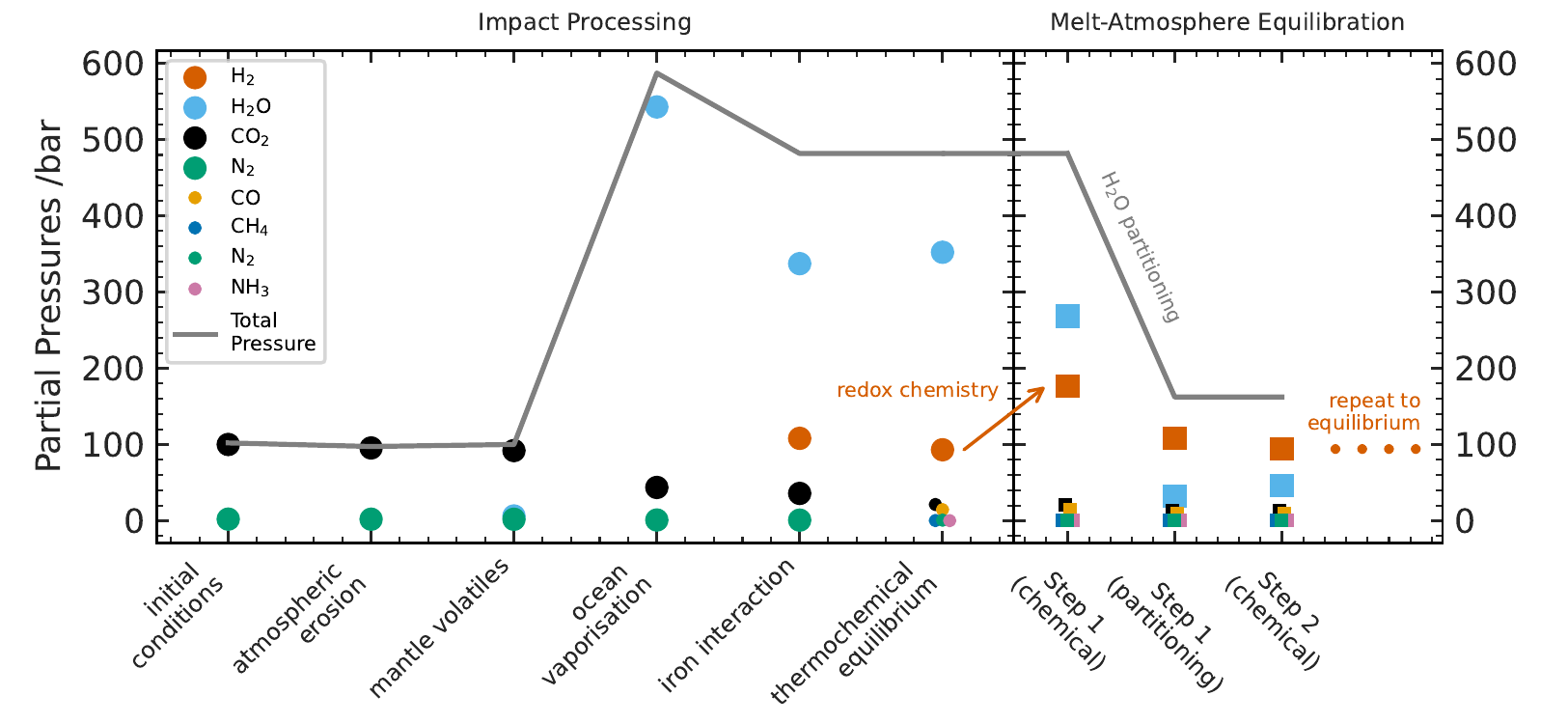}
    \caption{Partial pressures of atmospheric species at each step of impact processing and melt-atmosphere equilibration. Total atmospheric pressure at each stage is also shown. We note that several of the processes occurring during impact will occur over similar timescales (e.g., Figure \ref{fig:Timeline}), and our representation of these processes here is simply the ordering we have implemented within our calculations. Species other than \ce{H2O} and \ce{H2} are not involved in melt-atmosphere interactions. Melt-atmosphere interactions dominantly cause changes in the atmosphere during the first step (i.e., Step 1), meaning that the system is almost at equilibrium, and only minor changes occur in subsequent steps.}
    \label{fig:Walk-Through}
\end{figure*}

Here, we present a detailed view of how our early Earth system changes under impact processing and melt-atmosphere interactions (Figure \ref{fig:Walk-Through}).  We present these results under standard initial conditions ($M_i=2.0\times10^{22}\,\mathrm{kg}$, $v_i = 2\,v_\mathrm{esc}$, $\theta = 45^\circ$) and under the assumptions of Model 3A (iron accreted by the target interior is available to react with the melt phase). We present both the cases of a peridotite-like and a basalt-like melt phase composition. 

As a result of impact erosion, the pre-impact atmosphere ($100\,\mathrm{bars}$ \ce{CO2} and $2\,\mathrm{bars}$ \ce{N2}) loses $\sim 5\,\%$ of its mass. The net effect of adding volatiles from the vaporised target and impactor mantles, and vaporisation of the target's surface oceans, is to increase the atmosphere's mass by a factor of $\sim 5$, leading to a significant rise in surface pressure and an inflated atmosphere. $28\,\%$ of the impactor's iron inventory is made available to the reduce the atmosphere. The result is $345\,\mathrm{bars}$ of \ce{H2O} and $105\,\mathrm{bars}$ of \ce{H2}. The application of thermochemical equilibrium in the atmosphere sees \ce{H2O} increase to $360\,\mathrm{bars}$ and \ce{H2} decrease to $85\,\mathrm{bars}$. Correspondingly, the previous $36\,\mathrm{bars}$ of \ce{CO2} is halved to $18\,\mathrm{bars}$ by the production of $18\,\mathrm{bars}$ of \ce{CO} as the reducing power of the iron redistributes itself between the H-, C- and N-bearing molecules. At the post-impact stage, the total pressure in our atmosphere within the box model is $\sim485\,\mathrm{bars}$, and the atmospheric $f$O$_2$ is 2 log units below the FMQ buffer.

The impact produces a total melt phase mass of $6.3\times10^{23}\,\mathrm{kg}$ ( $\sim16\,\%$ of the target mantle). This includes rained out silicate vapours and \ce{FeO} produced by the oxidation of iron by the atmosphere, which account for $<10\,\%$ and $<1\,\%$ of the total mass respectively. $59\,\%$ of the impactor iron is accreted by the target interior and deposited into the melt phase. The impact-generated melt phase is not oxidising enough to oxidise all of the accreted iron, for neither the peridotitic nor basaltic melt compositions. As such, both melt phases are metal saturated at the post-impact stage. In the case of the peridotitic melt, the initial ferric-to-iron ratio (\ce{Fe^3+} / $\Sigma\ce{Fe}$) of $\sim5.0\,\%$ is reduced to $\sim0.7\,\%$ by the accreted impactor iron. For the basaltic melt, the initial $\sim16\,\%$ is also reduced to $\sim1\,\%$. 

Because the melt phase is more reducing than the atmosphere at the end of impact processing (i.e., the post-impact stage), the first step in melt-atmosphere interactions, which is a redox step, increases the abundance of \ce{H2} in the atmosphere and decreases the \ce{H2O} abundance. The second interactions step, which is a partitioning step, finds an atmosphere rich in \ce{H2O} in comparison to the relatively dry melt phase, and as such dissolves \ce{H2O} into the melt until equilibrium is achieved (Equation \ref{eqn:H2O Equilibrium Calc}). The two steps repeat, making smaller and smaller changes to the system, until both equilibria are satisfied simultaneously. 

Upon completion of the melt-atmosphere interactions (i.e., the post-equilibration stage), the atmosphere has shrunk in total pressure by a factor of $\sim3$, to $160\,\mathrm{bars}$. The draw down of \ce{H2O} into the melt phase is mostly responsible for this. Only $20\,\mathrm{bars}$ of \ce{H2O} remains in the atmosphere at equilibrium in the case of the peridotite, and only $40\,\mathrm{bars}$ with the basalt. The \ce{H2} partial pressure is now $125\,\mathrm{bars}$ in the case of the peridotite melt, and $100\,\mathrm{bars}$ for the basalt. This corresponds to molar increases in \ce{H2} by factors of $\sim3$ and $\sim2$, respectively. The post-equilibration atmospheres are thus \ce{H2}-dominated, with large components of \ce{H2O} and carbon-bearing species. The post-equilibration atmosphere and melt phase $f$O$_2$ are equal by definition, at half a log unit above metal saturation ($\Delta\mathrm{FMQ}=-5.0$) for the peridotite melt, and $\sim1.5$ log units above metal saturation ($\Delta\mathrm{FMQ}=-4.0$) for the basalt melt. The ferric-to-iron ratio ($\ce{Fe^3+}/\Sigma\ce{Fe}$) has also correspondingly increased.

%% file: sections/5.2_model_comparisons.tex
\subsection{The five models} \label{sec:Five Models}
The novel physics in this study are the distribution of the impactor iron within the system, and the interactions between the impact-generated melt phase and the post-impact atmosphere. We now focus on demonstrating the importance of these effects to considerations of early Earth. We compare the five models detailed in Figure \ref{fig:Graphical Abstract} (\textit{right}), which differ in how the impactor iron is distributed and whether or not the melt-atmosphere interactions are accounted for. In all cases, we use the standard values detailed in Table \ref{tab:Standard Parameters}. We consider both peridotitic and basaltic melt phases, representing our high- and low-fraction melting scenarios respectively. We further compare to the results of \citet{zahnle2020creation}, which our impact scenarios are very similar to in nature, and which we shall refer to henceforth as Z20.

\subsubsection{Model 1A (Fiducial) and Model 1B} \label{sec:Model 1}

Models 1A \& 1B calculate the state of the atmosphere in isolation from the target’s interior; there is no equilibration with the impact-generated melt phase. What we have previously termed the post-impact and post-equilibration states in this study are, therefore, identical to one another for the atmosphere. Z20 also make this assumption, but additionally assume that the impactor iron is made fully available to reduce the vaporised oceans. Model 1A and Z20 thus make the same assumptions in these respects. We hence term Model 1A the Fiducial Model, as it most closely represents previous work. In Model 1B, the iron inventory is distributed between the target atmosphere, target interior, and escaping the system (Section \ref{sec:Iron}). The lack of melt-atmosphere interactions means that the reducing power of iron distributed to the interior is never made available to the atmosphere. Model 1B thus demonstrates the effect of iron distribution in isolation from considerations of the impact-generated melt phase.
\begin{figure*}
    \centering
    \includegraphics[width=0.9\textwidth]{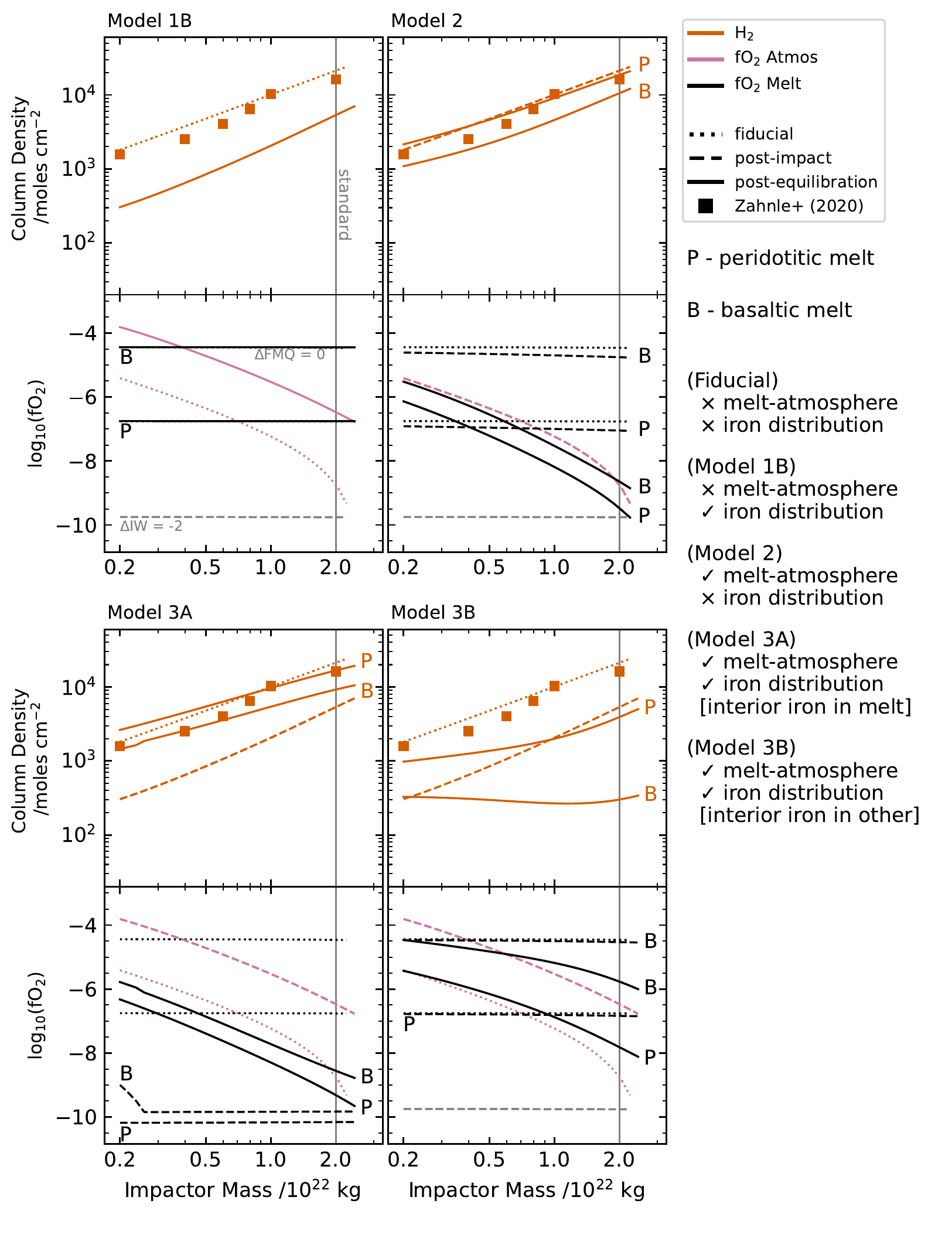}
    \caption{\ce{H2} abundances, atmospheric $f$O$_2$, and melt phase $f$O$_2$ as a function of impactor mass. Results for both the peridotitic (P) and basaltic (B) melt phases are shown. The results for the Fiducial Model (1A) are shown in all panels for comparison (dotted). Comparable results from \citet{zahnle2020creation} are also shown in all panels (squares). For each of the other models shown in Figure \ref{fig:Graphical Abstract} and detailed in Section \ref{sec:Five Models}, the post-impact (dashed) and post-equilibration (solid) results are shown. The standard values of Table \ref{tab:Standard Parameters} are used for all parameters except impactor mass in all cases. The standard impactor mass is indicated by the vertical line for each model.}
    \label{fig:Five Models}
\end{figure*}

All panels in Figure \ref{fig:Five Models} contain the results for the Fiducial Model (dotted lines). We observe high \ce{H2} abundances due to the impactor iron fully interacting with the atmosphere. The atmosphere is reduced relative to the FMQ buffer, and even drops below the IW buffer for large impactor masses. The $f$O$_2$ shown for the Fiducial melt is the $f$O$_2$ of a peridotitic or basaltic melt at generation (Section \ref{sec:Target}). All panels also show the results of Z20 for the same planetary initial conditions as our standard values\footnote{We ran models using the Z20 code available at \url{doi:10.5281/zenodo.3698264} with starting conditions of $p\ce{H2O}=500\,\mathrm{bars}$,  $p\ce{CO2}=100\,\mathrm{bars}$, and  $p\ce{N2}=2\,\mathrm{bars}$ (all other species set to $0\,\mathrm{bars}$), and extracted the atmospheric compositions at our system temperature of $1900\,\mathrm{K}$.} (squares).  The \ce{H2} abundances of the Fiducial Model and Z20 are highly alike, with a maximum relative difference of $40\,\%$. We see similar differences in the post-impact atmospheres (Figure \ref{fig:Appendix Zahnle}, Appendix \ref{sec:Appendix post-impact}), however, suggesting that these small differences are created during impact processing and then propagated forwards into our models' equilibrium states, rather than being a feature of melt-atmosphere interactions themselves. This is reassuring that the Fiducial Model provides a good basis for comparison between our models. The small differences between our calculations and those of Z20 most likely stem from the different thermochemical data being used (i.e., our \texttt{FastChem} calculations at the final step of impact processing, and the Z20 thermochemistry - see their Appendix B). 

In Model 1B (solid line in Figure \ref{fig:Five Models}, \textit{upper left}), the atmospheric abundance of \ce{H2} is substantially lower than in the Fiducial Model and in Z20 across the impactor mass range. The addition of iron distribution effects has removed reducing power from the atmosphere. Atmospheres in Model 1B are thus at least 1 log unit more oxidising than the Fiducial case.

\subsubsection{Model 2}\label{sec:Model 2}

Melt-atmosphere interactions are included in Model 2. The post-equilibration system is thus distinct from the post-impact system. Model 2, however, keeps the other assumption of the Fiducial Model, with all of the impactor iron being made available to reduce the target atmosphere during impact processing. Due to the identical initial treatment of iron accretion, the post-impact atmosphere in Model 2 (dashed line in Figure \ref{fig:Five Models}, \textit{upper right}) is identical to the Fiducial case. The melt phase, however, is slightly reduced in Model 2 relative to the Fiducial due to accretion of \ce{FeO} produced via oxidation of the impactor iron by \ce{H2O} in the atmosphere. 

In the case of the peridotitic melt, the post-impact atmosphere is more reducing than the melt for $M_i \gtrsim 10^{22}\,\mathrm{kg}$, and more oxidising than the melt for masses smaller than this. The relative difference between the atmosphere and melt phase $f$O$_2$, however, is less than 2 log units for all $M_i$. The resulting changes in \ce{H2} abundances between the post-impact and post-equilibration stages (i.e., changes caused by melt-atmosphere interactions only) are thus small. In the case of the basaltic melt, which is produced in a more oxidising state than the peridotitic melt, the post-impact atmosphere is more reduced than the melt phase for all $M_i$. As a result, the \ce{H2} abundance at melt-atmosphere equilibrium is lower than at post-impact for all $M_i$ by a factor of $\sim2$. Across both melt compositions, the equilibrium system becomes more reduced with increasing impactor mass, as more reducing power is introduced into the system via the larger impactor iron core. The melt phase always remains above metal saturation, however, meaning that there is never any metallic iron left in the melt pool at equilibrium.

Model 2 shows the effects of including melt-atmosphere interactions in isolation from considerations of the iron distribution. The changes in the atmosphere are relatively minor due to the similarities in starting oxygen fugacities of the melt and atmosphere, with maximum changes in \ce{H2} abundances only being a factor of 2. However, the $f$O$_2$ of the impact-generated melt phase can be substantially altered by the melt-atmosphere interactions.

\subsubsection{Model 3A}\label{sec:Model 3A}

The combined effects of impactor iron distribution and melt-atmosphere interactions are now presented. In Model 3A, we assume that all iron accreted by the target interior is available to reduce the impact-generated melt phase. This is an end member case, and we explore the alternative extreme in Model 3B. 

The proportion of iron made available to the atmosphere in Model 3A is the same as in Model 1B. The post-impact atmospheres are thus identical in both models. The post-impact melt phase of Model 3A is distinct from Model 1B, however, through its reduction by the impactor iron (Figure \ref{fig:Five Models}, \textit{lower left}). For the peridotite-like melt phase, sufficient iron is accreted by the melt phase such that it is metal saturated for all $M_i$. Basalt-like melt phases are also metal saturated, except for the smallest impactors of our considered mass range. 

The post-impact melt phases are more reducing than the atmospheres for both melt compositions, although the gap in redox state between the two reservoirs decreases with increasing impactor mass as the atmosphere becomes more reduced. Melt-atmosphere interactions thus serve to reduce the atmospheres, and increase atmospheric \ce{H2} abundances. This effect is more pronounced in the case of peridotitic melts due to these melts hosting more reducing power than the basaltic melts at the post-impact stage. The atmospheres equilibrated with the peridotitic melts thus have \ce{H2} abundances within a factor of $\sim2$ of the Fiducial Model. Atmospheres equilibrated with the basaltic melts, on the other hand, maintain \ce{H2} abundances lower than the Fiducial for all $M_i$. The slight inflection in the Z20 data with $M_i$ means that the two post-equilibration \ce{H2} abundances straddle the Z20 data.

The similarity in equilibrium states of Models 2, 3A, the Fiducial, and Z20 together demonstrate one of our major findings: as long as the impactor's iron inventory is accreted by either the melt phase or the atmosphere, and the two reservoirs are allowed to interact until they reach equilibrium, the whole of the iron's reducing power can affect the redox state of atmosphere. \ce{H2} abundances will thus be approximately the same as if all of the iron had been accreted by the atmosphere in the first place (e.g., Z20). Only the reducing power lost through iron remaining unaccreted during impact remains inaccessible to the atmosphere. While this is small for our $\theta_i = 45^\circ$ impacts, this loss could still be substantial in other impact scenarios.
\begin{figure*}
    \centering
    \includegraphics[width=0.85\textwidth]{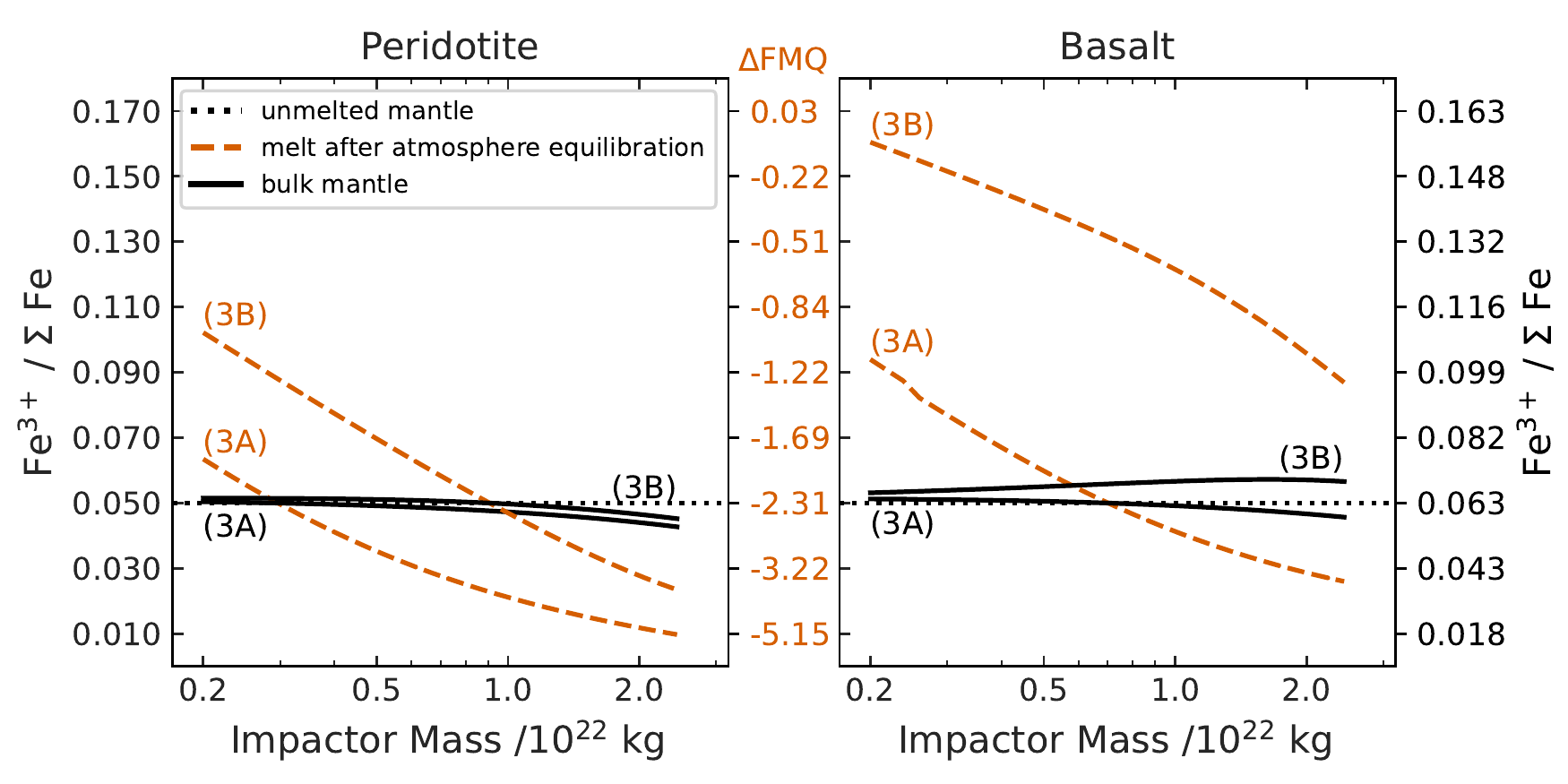}
    \caption{Ferric-to-iron ratio (\ce{Fe^3+}/$\sum\mathrm{Fe}$) as a function of impactor mass for the impact-generated melt phase at melt-atmosphere equilibrium (dashed), the remainder of the target mantle which is unmelted (dotted), and the combined bulk mantle (solid). The corresponding $f\!\mathrm{O}_2$ of the melt phase is also shown relative to the FMQ buffer (red). Results are shown for both Models 3A \& 3B.}
    \label{fig:Mantle Remixing}
\end{figure*}

\subsubsection{Model 3B}\label{sec:Model 3B}
Model 3B is the alternate end member case to Model 3A. Here, none of iron accreted by the target interior during impact is deposited into the impact-generated melt phase. Instead, iron is deposited into reservoirs inaccessible during the melt-atmosphere equilibration (e.g., large chemically inaccessible iron blobs which sink out of the melt phase). The post-impact melt phase $f$O$_2$ is thus close to the $f$O$_2$ at which it formed (Figure \ref{fig:Five Models}, \textit{lower right}), with the only change being the addition of \ce{FeO} from oxidation of the iron accreted by the atmosphere (there is less \ce{FeO} added than in Model 2 as now less iron is accreted by the atmosphere, hence the post-impact melt is more oxidised in Model 3B). There is no change in the post-impact atmosphere from Model 3A. 

For the peridotitic melts, the post-impact melt phase is more reducing than the atmosphere for all $M_i$. We thus see melt-atmosphere interactions leading to post-equilibration \ce{H2} abundances greater than post-impact, but only for impacts with $M_i \lesssim 10^{22}\,\mathrm{kg}$. The equilibrium atmospheres under impacts with $M_i \gtrsim 10^{22}\,\mathrm{kg}$, however, see post-equilibration \ce{H2} abundances lower than post-impact, despite interacting with melt phases more reducing than themselves. The cause of this counter intuitive result is the partitioning of \ce{H2O} between the atmosphere and melt phase during melt-atmosphere interactions. The dissolution of \ce{H2O} into the melt allows for more oxidising of the atmosphere before equilibrium is achieved than would otherwise be possible in its absence. For the basaltic melts, the post-impact melt phase is more oxidising than the atmosphere for most $M_i$. As such, the post-equilibration \ce{H2} abundances are lower than the post-impact for all $M_i$, particularly at larger impactor masses, where the logarithmically increasing melt mass enables greater dissolution of \ce{H2O} into the melt phase and intrinsically provides more oxidising power to the system. The increase in available oxidising power as a result of increasing melt mass is faster as a function of $M_i$ than the increase in available reducing power as a result of the larger impactor core. It is this effect that is responsible for the curvature in equilibrium \ce{H2} abundances seen as a function of impactor mass for both melt phase compositions (Figure \ref{fig:Five Models}, \textit{lower right}).

The post-equilibration atmospheres of Model 3B are substantially more oxidised than those of Model 3A and Z20. This demonstrates another of our major findings: the fraction of iron accreted by the target interior that is available to reduce the impact-generated melt phase is key in understanding the potential reducing power of large impacts. \ce{H2} abundances can be around an order of magnitude lower as a result of switching from Model 3A to 3B, for both melt phase compositions. As with the iron that escapes the system, iron that cannot chemically interact with the melt phase simply represents a loss of reducing power from the system. For our $\theta_i = 45^\circ$ impacts, the large fractions of the impactor iron being accreted by the target interior mean that transitioning from Model 3A to Model 3B results in a much greater loss of reducing power from the system than that lost from escaping iron. However, for more obtuse impacts, where a substantially greater portion of the impactor iron escapes the system, escaping iron will likely represent a greater loss of reducing power.

%% file: sections/5.3_remixing.tex
\subsection{Melt phase and solid mantle redox states}\label{sec:Mantle Remixing}
At melt-atmosphere equilibrium, the melt phase represents a distinct reservoir within the target mantle. The difference in ferric-to-rion ratio (\ce{Fe^3+}/$\sum\mathrm{Fe}$) and $f$O$_2$ between the equilibrated melt phase (dashed lines, Figure \ref{fig:Mantle Remixing}) and the unmelted solid mantle (dotted lines) depends on whether we use Model 3A or 3B. This is expected due to the generally more oxidising state of the equilibrated melt-atmosphere system in Model 3B. 

For Model 3A, melt phases formed by our largest impactors have ferric-to-iron ratios around $2-5$ times lower than the unmelted mantle, depending on the melt composition (Figure \ref{fig:Mantle Remixing}). Melts formed by our smaller impactors can have \ce{Fe^3+}/$\sum\mathrm{Fe}$ up to $\sim1.5$ times greater than the solid mantle. For Model 3B, the equilibrated melt phase is only more reducing than the unmelted mantle for peridotitic melt phases at large impactor masses ($M_i \gtrsim 10^{22}\,\mathrm{kg}$). In most cases under Model 3B, therefore, the melt phase is a reservoir of oxidising power within the mantle, with \ce{Fe^3+}/$\sum\mathrm{Fe}$ up to $\sim2.5$ times greater than the unmelted mantle. The solidified melt phase can thus represent an oxidising or reducing redox heterogeneity within the planet's mantle, depending on how the impactor iron is accreted by the melt.

Due to the relatively small mass of the melt in comparison to the unmelted mantle, the bulk redox state of the mantle does not vary substantially due to the presence of the equilibrated melt phase (solid lines, Figure \ref{fig:Mantle Remixing}). 

%% file: sections/6_discussion.tex
\section{Discussion}\label{sec:Discussion}
The five models we have presented in Section \ref{sec:Five Models} demonstrate that the melt-atmosphere system is more oxidising at equilibrium when: 
\begin{enumerate}
    \item{A greater portion of the impactor iron inventory is not accreted by the target;}
    
    \item{A greater portion of the impactor iron inventory that is accreted by the interior is accreted in a way that does not allow it to chemically interact with the impact-generated melt phase; or}
     
    \item{The melt phase composition is more basalt-like than peridotite-like.}
\end{enumerate}  
Our current suite of simulations characterise well the non-accretion of iron. However, improved simulations are required to determine where the iron ends up within the target interior, and hence where the balance exists between the two end-member cases of Models 3A \& 3B. Such improved simulations would also aid in determining the balance between peridotitic and basaltic melt phase composition by providing more accurate melt fraction data, although we re-iterate that the peridotitic melts will likely be more representative based on our current simulations. 

There are several implications of the early Earth's atmosphere being more oxidising in the aftermath of large impacts. The presence of a mantle reservoir with a distinct redox state also has consequences during the planet's evolution, if such impacts took place. In our discussion of these effects, we make no predictions as to whether large impacts are better characterised by Model 3A or 3B, instead accounting for both sets of results where appropriate. However, based on the melt fraction data generated by our simulations (see Section \ref{sec:Melt Composition}), we do consider the case of the peridotitic melt phase as being a closer to what would be found in impacts.  

\input{sections/6.1_global_conditions}

\input{sections/6.2_locally_reduced}

\input{sections/6.3_iron}

%% file: sections/6.1_global_conditions.tex
\subsection{Global conditions in the aftermath of impact}\label{sec:Discussion Evolution}
\subsubsection{Thermochemistry in the atmosphere}
Having more oxidising atmospheres in the aftermath of large impacts does not necessarily prevent reduced species forming in sufficient quantities for prebiotic chemical pathways. In the end-member case of Model 3B, our most oxidising case in terms of the atmosphere, there are still tens of bars of \ce{H2} remaining in the atmosphere at melt-atmosphere equilibrium for our standard impactor mass ($2.0\times10^{22}\,\mathrm{kg}$). A coupled climate-chemistry model, such as that of Z20, is beyond the scope of this study. However, simplified calculations of atmospheric cooling under thermochemical equilibrium can give us an indication of what we might expect as a result of lower \ce{H2} abundances. Using \texttt{FastChem}, we cooled the atmospheres of Models 3A, 3B, and the Fiducial from the initial post-impact $1900\,\mathrm{K}$ to just above nitrogen chemistry quenching\footnote{We stop the cooling at the nitrogen chemistry quench point to avoid carrying out nitrogen gas phase chemistry that should have quenched, a process that the thermochemical equilibrium solver does not account for.} ($\sim1100\,\mathrm{K}$, Z20). As expected, due to similar abundances of \ce{H2} at melt-atmosphere equilibrium, the differences between Model 3A and the Fiducial Model are small. At the standard impactor mass, Model 3A hosts $95\,\%$ of the Fiducial Model's \ce{CH4} abundance, and $65\,\%$ of its \ce{NH3} abundance. Atmospheres in Model 3B, however, host lower abundances of the reduced species. At the standard impactor mass, the \ce{CH4} abundance is only $35\,\%$ of the Fiducial abundance, and \ce{NH3} sits at $15\,\%$. While low, we suggest that these end-member abundances might yet be sufficient for \ce{HCN} formation via photochemistry to occur in quantities suitable for prebiotic chemical pathways. 

\subsubsection{Surface temperatures}
High surface temperatures are a known issue in using large impacts to provide reducing global conditions for prebiotic chemistry. In the Z20 scenario, for example, after condensation of the steam atmosphere and before the \ce{H2} escapes the Earth over 10,000-year timescales, high abundances of \ce{H2} would lead to a strong greenhouse effect through collision-induced absorption (e.g., \citealt{pierrehumbert2011hydrogen, wordsworth2013hydrogen}). This, combined with some additional \ce{CH4}-driven greenhouse heating, would keep surface temperatures far above those suitable for prebiotic chemical pathways (e.g., \citealt{mansy2008thermostability, attwater2013ice, rimmer2018origin}) long after the impact. While waiting for the system to cool, the surface environment is being re-oxidised by outgassing from the planet's mantle, meaning that the window of opportunity narrows for prebiotic chemistry that relies on globally reducing surface conditions \citep{benner2020when}.

Models 3A \& 3B differ from this scenario in two relevant ways. Firstly, after liquid water has re-condensed on Earth's surface, the more oxidising atmospheres (Model 3B in particular) will experience diminished greenhouse heating in comparison to the Fiducial Model as a result of lower \ce{H2} and \ce{CH4} abundances. This goes some way to fixing the surface temperature issues. However, there is a necessary balance between this fix and yet still having sufficient quantities of reducing species to enable prebiotic pathways. The reducing conditions of Model 3A, the Fiducial Model, and Z20 may keep the surface too hot. However, moving towards Model 3B may more readily satisfy this balance. The second difference in the atmosphere's thermal behaviour is caused by the melt phase acting as a heat source at the base of the atmosphere. This extends the lifetime of the steam atmosphere, whose cooling rate is limited at the runaway greenhouse limit. Dissolution of \ce{H2O} into the melt phase would temporarily diminish the atmospheric greenhouse effect, but upon crystallisation of the melt phase, \ce{H2O} and \ce{CO2} will be outgassed into the atmosphere, and the greenhouse will resume \citep{elkins2008linked, lebrun2013thermal, nikolaou2019factors}. As with the \ce{H2} escape, delaying the time after impact at which we reach clement surface conditions increases the opportunity for the Earth's mantle to re-oxidise the surface, and narrows the window of opportunity for prebiotic chemistry.

\subsubsection{Highly siderophilic elements}
The accretion of impactor material to the impact-generated melt affects the redox state of the local mantle region (Section \ref{sec:Discussion Local}) as well as the reservoir's trace element and isotopic signatures (e.g., \citealt{marchi2018heterogeneous, maas2021fate}). In particular, anomalously high relative abundances of HSEs, and anomalously low $\epsilon\ce{^182 W}$, could be detectable in ancient igneous rocks stemming from this region if such impacts took place on early Earth. This effect could, however, be masked by efficient global mixing of Earth's mantle (e.g., \citealt{paquet2021effective}).

The differences in iron interactions between Models 3A \& 3B can affect the abundances of HSEs deposited in the mantle during our large impact scenarios. In Model 3A, the iron fully mixing with the melt means that HSEs would be added to the upper mantle in roughly chondritic proportions. In Model 3B, the lack of any interaction between the iron and the melt results in the HSE abundance of the upper mantle only being perturbed by the addition of the silicate component of the impactor. This impactor silicate-derived HSE signal will likely be small due to the relative masses of the added material (Earth mantle mass is $\sim250$ times more massive than our largest impactor's mantle mass). With large impacts likely to reflect outcomes somewhere in between our two models, there is scope for accretion of HSEs in non-chondritic proportions, something which does not match observations of Earth's mantle (e.g., \citealt{bottke2010stochastic, rubie2015accretion, rubie2016highly}) and which may weaken the evidence that such large impacts took place on early Earth. The relative losses of core and mantle material from the impactor, and the initial distribution of HSEs within the impactor, are key in resolving this. HSEs could still be accreted by the target mantle in chondritic proportions, despite loss of impactor core material through the accretion processes leading to our Model 3B, if fractional losses of core and mantle material occur in particular ratios, or if the impactor did not host chondritic proportions of HSEs. Further analysis on the relative losses of core and mantle impactor material are required.

%% file: sections/6.2_locally_reduced.tex
\subsection{Formation of a locally reduced melt phase}\label{sec:Discussion Local}
\subsubsection{Long lived Fe-redox heterogeneity in the mantle}
The crystallised impact-generated melt phase exists as a distinct redox reservoir within the target mantle. The bulk Fe-redox state of the target mantle does not vary substantially in the presence of this reservoir (Section \ref{sec:Mantle Remixing}), even at the extremes of our impactor mass range. The variation is especially small when compared to uncertainties in Earth's Hadean mantle $f$O$_2$ (e.g., \citealt{cavosie2006correlated, trail2011oxidation}). However, the difference in \ce{Fe^3+}/$\sum\mathrm{Fe}$ between the crystallised melt phase and the remainder of the target mantle can be substantial (Figure \ref{fig:Mantle Remixing}).

Melt phase redox reservoirs could, therefore, be responsible for heterogeneities within Earth's early mantle if large impacts took place, and thus could affect the distribution of measured mantle redox states. Indeed, igneous zircons record some tentative evidence for reducing conditions in the Hadean \citep{yang2014relatively}, in line with our findings from larger impactors. The melt pools formed from smaller impactors tend to be more oxidising than the the bulk mantle (Figure \ref{fig:Mantle Remixing}), and so would stretch the observed distribution of early mantle redox states towards more oxidising conditions. The substantially smaller volumes of melt produced by the smaller mass impactors would limit the likelihood of sampling such regions, but on the other hand we would expect more of these impactors to have hit the early Earth. The reducing effects of larger impacts should be more evident in measurements. Future estimates of Earth's Hadean/Archean mantle $f$O$_2$ should be investigated for evidence of these transient reducing domains.

\newpage\subsubsection{Re-emergence of reducing power at the planet's surface}
On long timescales, the materials of the crystallised impact-generated melt phase can be recycled back to the planet surface. For larger impacts, especially around our standard impactor mass, the crystallised melt phase will be reducing relative to the bulk mantle (Figure \ref{fig:Mantle Remixing}). The re-emergence of this material thus creates the possibility of locally reducing conditions at the planet's surface. Volcanism stemming from the locally reduced mantle will degas a reducing ensemble of volatiles (e.g., \citealt{liggins2020can}), extending the period over which reduced species are available at the planet surface. This is useful for some suggested pathways for prebiotic chemistry, which require the presence of both oxidised and reduced species (e.g., \citealt{benner2020when}). Subsurface hydrothermal vents are another way in which reducing mantle volcanism could affect the planet surface, with \citet{rimmer2019origin} demonstrating that reduced species outgassed at these vents, upon reaction with subsurface water, can produce useful quantities of feedstock molecules for prebiotic chemistry.

We thus suggest that not only can large impacts lead to a global environment amenable to hosting the origin of life, such as the photochemical production of \ce{HCN} (e.g., \citealt{benner2020when,zahnle2020creation}), but they can also give rise to plausible scenarios relying on locally reduced conditions for potentially 10s of Myr post-impact.

%% file: sections/6.3_iron.tex
\subsection{Comments on iron distributions}\label{sec:Discussion Iron}
The differences between the equilibrium states of Models 3A and 3B are substantial (Figure \ref{fig:Five Models}), and thus so are the potential consequences for Earth's evolution in the time after large impact events (Sections \ref{sec:Discussion Evolution} \& \ref{sec:Discussion Local}). Where the balance sits between the two models is thus a vital question. 

Iron simply not being accreted by the melt phase is one way to remove its reducing power from the melt-atmosphere interacting system. Firstly, if the impact is sufficiently large, the timescale of isostatic readjustment will be less than the timescale of magma solidification, and buoyant melt will ascend and spread over the surface \citep{tonks1993magma, reese2006fluid}. The iron is more dense than this evacuating magma, and could be left behind during the magma ascension, and buried in the solid mantle. The impacts considered in this study are likely not sufficiently energetic for this process to occur substantially \citep{reese2006fluid}, however, the effect will become stronger with more head-on collisions. Secondly, simulations of large planetesimal impacts (e.g., \citealt{genda2017terrestrial, citron2022large}) suggest that portions of the projectile core may remain bound as large coherent fragments of iron. Large iron fragments may be sequestered directly into the solid mantle, and even the planet core, if they are quickly buried in the collapse of the transient crater cavity \citep{tonks1992core} or if they are not significantly disrupted into smaller fragments through pancaking and stretching \citep{kendall2016differentiated}. This process is a likely way through which loss of reducing power from the melt phase can occur. However, it is difficult to assess in purely hydrodynamic simulations of such impacts, as the behaviour depends on the thermal profile and rheology of the target interior, as well as a suite of microphysical processes, and is not well captured. Future simulations should attempt to capture such behaviours.

Iron being accreted by the impact-generated melt phase, but then not being able to take part in melt-atmosphere interactions is another way to remove reducing power from the system. In our calculations, we assume that the melt phase is one large homogeneous body, well-mixed by vigorous convection. However, the completeness of melting generally decreases with distance from the impact point \citep{tonks1993magma}, and the higher viscosity of partial melt in comparison to complete melt can cause the melt phase to become stratified, with two distinct layers  \citep{abe1995basic,reese2006fluid}. The fully molten layer with lower viscosity would be able to convect and  fully interact with the atmosphere. The layer consisting of melt-crystal slurry, however, would have higher viscosity and would less efficiently communicate with the atmosphere, and reducing power accreted to this reservoir would be at least partially lost to the surface environment.

In our simulations, we find that only $65$-$75\,\%$ of our impact-generated melt phase is fully molten (i.e., with melt fraction $\sim100\,\%$). However, $\sim90\,\%$ of our melt across the simulations reaches melt fractions greater than $40\,\%$, where silicate melts experience a sharp drop-off in viscosity \citep{abe1995basic}. We thus expect an overall high mobility of the impact-generated melt phase, and we suggest that stratification would have a limited effect on the system equilibrium state. Conversely, if future simulations show that the impactor iron has a high propensity to be buried in mantle regions with low-fraction melting, restricted melt phase mobility may become a more important effect. In this case, impactor iron can become trapped in the immobile layer, removing reducing power from the melt-atmosphere system \citep{solomatov2007magma}.

%% file: sections/7_conclusion.tex
\newpage\section{Conclusions}\label{sec:Conclusions}
Large impacts have been suggested as promising scenarios under which reduced species relevant to prebiotic chemistry can form \citep{benner2020when, zahnle2020creation}. Here, we demonstrate the importance of accounting for the distribution of the impactor's iron inventory during accretion by the planet in such scenarios, as well as accounting for chemical and physical interactions between the atmosphere and the impact-generated melt phase. The combination of these processes is key to understanding the equilibrium state of the system, and how much of the impactor's reducing power is accessible to generate reduced atmospheric species.

We find that as long as the impactor iron's reducing power is accreted by either the atmosphere or the vigorously convecting melt phase, melt-atmosphere interactions ensure that the reducing power is able to affect the atmospheric redox state at equilibrium. Reducing power can be lost from the system, and thus a more oxidising atmosphere can be produced, when iron escapes the system during impact or is accreted by the melt phase in such a way (e.g., large chemically inaccessible blobs) that it is unable to reduce the melt and hence take part in melt-atmosphere interactions. The system will also be more oxidising at melt-atmosphere equilibrium in the case that a greater portion of the impact-generated melt phase is formed as low-fraction melt, which is more oxidising than high-fraction melt at formation, and is less likely to be able to convect to the surface of the melt pool to take part in atmosphere interactions. The magnitude of these oxidising effects is greater for larger impactor masses, as the effects are magnified by the logarithmically increasing melt mass produced during impact.

Improved impact simulations are required to determine more accurate distributions of impactor iron within the target mantle. Future simulations should seek to include appropriate elastic-plastic interior rheologies, a feature which should also improve the assessment of high- and low-fraction melt distributions. Such simulations could also help to assess the abundances of highly siderophile elements that we would expect to be accreted by the target mantle, and whether these abundances match the chondritic relative proportions observed in Earth's mantle today. Purely hydrodynamic simulations may not be suitable for such tasks.

Large impacts could have led to the creation of reduced environments on early Earth with species relevant to prebiotic chemistry. However, we suggest that these environments could have been oxidising enough that issues associated with a reduced greenhouse climate would have been suppressed. Further, the creation of a locally reduced reservoir within the mantle had the potential to contribute to surface conditions for prebiotic chemistry over geologic time, and provide a record of such impacts having occurred.

%% file: appendices/A_GADGET.tex
\section{GADGET2 SPH simulations}\label{sec:Appendix GADGET}
\renewcommand{\theequation}{A\arabic{equation}}
\renewcommand{\thetable}{A\arabic{table}}
GADGET2 SPH \citep{springel2005cosmological} was modified to utilise the tabulated equations of state to model planetary collisions \citep{marcus2009collisional,marcus2011role}. The thermodynamic behaviour of the materials is described by equations of state generated by the analytic equation of state program (ANEOS; \citealt{thompson_improvements_1974,melosh2007hydrocode,thompson2019MANEOS,stewart2019ANEOS}) using the modifications included in M-ANEOS v1 \citep{thompson2019MANEOS}. The mantle material of the target and impactors is represented by an equation of state for forsterite \citep{stewart2019equation}, while the core material for each of the bodies is represented by an equation of state for \ce{Fe_{85}Si_{15}} iron alloy \citep{stewart2020equation}. The effects of material strength were not included. 

The impactors and Earth were initialised using WoMa (World Maker) \citep{ruiz2021effect}. The Earth was initialised with an isentropic thermal profile for the solid material with a specific entropy of $1800\,\mathrm{J\,kg}^{-1}\mathrm{\,K}^{-1}$ for iron and $2810\,\mathrm{J\,kg}^{-1}\mathrm{\,K}^{-1}$ for forsterite, corresponding to a mantle potential temperature of $\sim1900\,\mathrm{K}$. Impactors were initialised with an adiabatic temperature profile corresponding to the same surface temperature. Prior to the full impact simulations, all bodies were run in isolation for up to 12 hours of relaxation time, so that the rms velocity of the SPH particles converged to a small initial value. Due to computational constraints, simulations were limited to between $5\times10^5$ and $3\times10^6$ total SPH particles, so that the impactors consisted of at least 4000 particles regardless of their size. The simulations utilize indivisible particles $\sim150-250\,\mathrm{km}$ in diameter, and therefore do not resolve any crust or atmospheric mixing.

The melt mass is determined by examining the phase of each SPH particle using the ANEOS phase boundaries in Pressure-Entropy space. The mass fraction of liquid for mixed-phase states is determined using the lever rule. The melt mass is taken as the sum of all liquid material including the melt fraction of partially molten particles in the post-impact Earth and the bound atmosphere. For the purposes of this calculation we also count supercritical fluid particles as completely molten. For further details see \citet{citron2021large}.
\begin{table}
    \centering
    \caption{GADGET2 simulation results for 48 different impact scenarios.}
    \label{tab:Impactor Sims}
    \begin{tabular}{ccccccc}
        \toprule
        $M_i$ /kg  &  $v_i$ /$v_\mathrm{esc}$  &  $\theta_i$  &  $M_\mathrm{melt}$ /kg  &  $X_\mathrm{int}$  &  $X_\mathrm{atm}$  &  $X_\mathrm{ejec}$  \\
        \midrule
        $7.167\times10^{21}$  &  1.1  &  00  &  $3.866\times10^{22}$  &  1.000  &  0.000  &  0.000  \\
        \quad                 &  1.1  &  30  &  $4.400\times10^{22}$  &  1.000  &  0.000  &  0.000  \\
        \quad                 &  1.1  &  45  &  $4.613\times10^{22}$  &  0.997  &  0.003  &  0.000  \\
        \quad                 &  1.1  &  60  &  $2.510\times10^{22}$  &  0.436  &  0.196  &  0.368  \\
        \quad                 &  1.5  &  00  &  $7.479\times10^{22}$  &  1.000  &  0.000  &  0.000  \\
        \quad                 &  1.5  &  30  &  $8.010\times10^{22}$  &  1.000  &  0.000  &  0.000  \\
        \quad                 &  1.5  &  45  &  $7.588\times10^{22}$  &  0.909  &  0.091  &  0.001  \\
        \quad                 &  1.5  &  60  &  $2.996\times10^{22}$  &  0.218  &  0.177  &  0.605  \\
        \quad                 &  2.0  &  00  &  $1.567\times10^{23}$  &  0.995  &  0.005  &  0.000  \\
        \quad                 &  2.0  &  30  &  $1.434\times10^{23}$  &  0.991  &  0.009  &  0.000  \\
        \quad                 &  2.0  &  45  &  $1.213\times10^{23}$  &  0.791  &  0.195  &  0.015  \\
        \quad                 &  2.0  &  60  &  $3.533\times10^{22}$  &  0.064  &  0.088  &  0.848  \\
        \midrule
        $1.792\times10^{22}$  &  1.1  &  00  &  $1.048\times10^{23}$  &  1.000  &  0.000  &  0.000  \\
        \quad                 &  1.1  &  30  &  $1.114\times10^{23}$  &  1.000  &  0.000  &  0.000  \\
        \quad                 &  1.1  &  45  &  $1.121\times10^{23}$  &  0.978  &  0.022  &  0.000  \\
        \quad                 &  1.1  &  60  &  $5.763\times10^{22}$  &  0.355  &  0.191  &  0.454  \\
        \quad                 &  1.5  &  00  &  $2.050\times10^{23}$  &  1.000  &  0.000  &  0.000  \\
        \quad                 &  1.5  &  30  &  $1.949\times10^{23}$  &  1.000  &  0.000  &  0.000  \\
        \quad                 &  1.5  &  45  &  $1.686\times10^{23}$  &  0.810  &  0.174  &  0.016  \\
        \quad                 &  1.5  &  60  &  $5.070\times10^{22}$  &  0.077  &  0.124  &  0.799  \\
        \quad                 &  2.0  &  00  &  $3.691\times10^{23}$  &  1.000  &  0.000  &  0.000  \\
        \quad                 &  2.0  &  30  &  $3.346\times10^{23}$  &  0.998  &  0.002  &  0.000  \\
        \quad                 &  2.0  &  45  &  $2.518\times10^{23}$  &  0.596  &  0.287  &  0.116  \\
        \quad                 &  2.0  &  60  &  $5.398\times10^{22}$  &  0.009  &  0.030  &  0.961  \\
        \midrule
        $3.583\times10^{22}$  &  1.1  &  00  &  $2.268\times10^{23}$  &  1.000  &  0.000  &  0.000  \\
        \quad                 &  1.1  &  30  &  $2.291\times10^{23}$  &  1.000  &  0.000  &  0.000  \\
        \quad                 &  1.1  &  45  &  $2.132\times10^{23}$  &  0.934  &  0.066  &  0.000  \\
        \quad                 &  1.1  &  60  &  $1.053\times10^{23}$  &  0.273  &  0.164  &  0.564  \\
        \quad                 &  1.5  &  00  &  $4.144\times10^{23}$  &  1.000  &  0.000  &  0.000  \\
        \quad                 &  1.5  &  30  &  $3.774\times10^{23}$  &  1.000  &  0.000  &  0.000  \\
        \quad                 &  1.5  &  45  &  $2.908\times10^{23}$  &  0.684  &  0.240  &  0.076  \\
        \quad                 &  1.5  &  60  &  $7.199\times10^{22}$  &  0.023  &  0.053  &  0.924  \\
        \quad                 &  2.0  &  00  &  $6.893\times10^{23}$  &  1.000  &  0.000  &  0.000  \\
        \quad                 &  2.0  &  30  &  $6.158\times10^{23}$  &  0.990  &  0.008  &  0.001  \\
        \quad                 &  2.0  &  45  &  $4.145\times10^{23}$  &  0.420  &  0.299  &  0.281  \\
        \quad                 &  2.0  &  60  &  $7.370\times10^{22}$  &  0.000  &  0.004  &  0.996  \\
        \midrule
        $7.167\times10^{22}$  &  1.1  &  00  &  $4.733\times10^{23}$  &  1.000  &  0.000  &  0.000  \\
        \quad                 &  1.1  &  30  &  $4.229\times10^{23}$  &  1.000  &  0.000  &  0.000  \\
        \quad                 &  1.1  &  45  &  $3.740\times10^{23}$  &  0.874  &  0.112  &  0.014  \\
        \quad                 &  1.1  &  60  &  $1.707\times10^{23}$  &  0.165  &  0.166  &  0.669  \\
        \quad                 &  1.5  &  00  &  $7.803\times10^{23}$  &  1.000  &  0.000  &  0.000  \\
        \quad                 &  1.5  &  30  &  $6.584\times10^{23}$  &  0.998  &  0.002  &  0.000  \\
        \quad                 &  1.5  &  45  &  $4.664\times10^{23}$  &  0.557  &  0.212  &  0.231  \\
        \quad                 &  1.5  &  60  &  $1.268\times10^{23}$  &  0.027  &  0.047  &  0.925  \\
        \quad                 &  2.0  &  00  &  $1.225\times10^{24}$  &  0.996  &  0.004  &  0.000  \\
        \quad                 &  2.0  &  30  &  $9.516\times10^{23}$  &  0.958  &  0.039  &  0.003  \\
        \quad                 &  2.0  &  45  &  $6.169\times10^{23}$  &  0.300  &  0.173  &  0.527  \\
        \quad                 &  2.0  &  60  &  $1.177\times10^{23}$  &  0.000  &  0.000  &  1.000  \\
        \bottomrule
    \end{tabular}
\end{table}

%% file: appendices/B_simultaneous.tex
\section{Formulation of simultaneous redox chemistry}\label{sec:Appendix Simultaneous}
\renewcommand{\theequation}{B\arabic{equation}}
\renewcommand{\thetable}{B\arabic{table}}
\subsection{Simultaneous oxidation or reduction of a peridotitic melt}
Through combining the prescriptions of \citet{sossi2020redox} and \citet{frost1991introduction}, the relative abundances of \ce{Fe2O3}, \ce{FeO}, and \ce{Fe} in a magma containing all three species can be written as,
\begin{equation}
    \log_{10}\left(\frac{X_\mathrm{\ce{Fe^3+}}}{X_\mathrm{\ce{Fe^2+}}} \right) = -0.504\,\log_{10}\left(\frac{X_\mathrm{\ce{Fe^0}}}{X_\mathrm{\ce{Fe^2+}}} \right) - 1.530~~,
\end{equation}
from which we can express,
\begin{equation}
    X_\mathrm{\ce{Fe^3+}} = 10^{-1.530}\,X_\mathrm{Fe^0}^{~~-0.504} \,X_\mathrm{\ce{Fe^{2+}}}^{~~1.504}~~,
\end{equation}
were $X_\mathrm{\ce{Fe^3+}}$ and $X_\mathrm{\ce{Fe^2+}}$ are the molar fractions of ferric and ferrous iron in the silicate melt phase, and $X_\mathrm{\ce{Fe^0}}$ is the purity of iron in the metal phase, which we take as 0.98. We can then express the changes in these relative abundances as
\begin{equation}
    \begin{split}
        \Delta X_\mathrm{\ce{Fe^3+}}  &= 10^{-1.530}\,X_\mathrm{Fe^0}^{~~-0.504} \left( 1.504\,X_\mathrm{\ce{Fe^{2+}}}^{~~0.504}\,\Delta X_\mathrm{\ce{Fe^{2+}}}\right) \\
        &= 1.504\,\left(\frac{X_\mathrm{\ce{Fe^3+}}}{X_\mathrm{\ce{Fe^2+}}}\right)\,\Delta X_\mathrm{\ce{Fe^2+}}~~.
    \end{split}
    \label{eqn:Combined Param}
\end{equation}
To convert these changes in molar fractions to changes in moles, we must account for the change in melt phase moles that occurs during the chemical reactions, 
\begin{equation}
    \Delta X_\mathrm{\ce{Fe^3+}} = \frac{n_\mathrm{\ce{Fe2O3}}^\prime}{2\,N^\prime} - \frac{n_\mathrm{\ce{Fe2O3}}}{2\,N}~~,
    \label{eqn:X_fe_3}
\end{equation}
where $N$ and $N^\prime$ are the total number of moles in the melt phase before and after the reactions respectively, and $n_\mathrm{\ce{Fe2O3}}$ is the number of moles of \ce{Fe2O3} in the melt phase. The factors of $1/2$ stem from there being twice as many moles of \ce{Fe^3+} than moles of \ce{Fe2O3} in the melt (i.e., $2\,n_\mathrm{\ce{Fe^3+}} = n_\mathrm{\ce{Fe2O3}}$). Similarly, 
\begin{equation}
    \Delta X_\mathrm{\ce{Fe^2+}} = \frac{n_\mathrm{\ce{FeO}}^\prime}{N^\prime} - \frac{n_\mathrm{\ce{FeO}}}{N}~~.
\end{equation}
We can express the total moles before and after the reactions as
\begin{equation}
    N^\prime = N + \Delta N = N + \Delta n_\mathrm{\ce{Fe2O3}} + \Delta n_\mathrm{\ce{FeO}}~~.
\end{equation}
We can then express Equation \ref{eqn:X_fe_3} as
\begin{equation}
    \Delta X_\mathrm{\ce{Fe^3+}} = \frac{N\,\Delta n_\mathrm{\ce{Fe2O3}} - n_\mathrm{\ce{Fe2O3}}\left(\Delta n_\mathrm{\ce{Fe2O3}} + \Delta n_\mathrm{\ce{FeO}}\right)}{2N \left(N + \Delta n_\mathrm{\ce{Fe2O3}} + \Delta n_\mathrm{\ce{FeO}} \right)}~~,
    \label{eqn:Delta X_fe_3}
\end{equation}
and similarly, 
\begin{equation}
    \Delta X_\mathrm{\ce{Fe^2+}} = \frac{N\,\Delta n_\mathrm{\ce{FeO}} - n_\mathrm{\ce{FeO}}\left(\Delta n_\mathrm{\ce{Fe2O3}} + \Delta n_\mathrm{\ce{FeO}}\right)}{N \left(N + \Delta n_\mathrm{\ce{Fe2O3}} + \Delta n_\mathrm{\ce{FeO}} \right)}~~.
    \label{eqn:Delta X_fe_2}
\end{equation}
Combining Equations \ref{eqn:Combined Param}, \ref{eqn:Delta X_fe_3} \& \ref{eqn:Delta X_fe_2}, we can express the total change in moles of \ce{FeO} across both chemical reactions (\ref{eqn:FMQ Reaction Appendix} \& \ref{eqn:IW Reaction Appendix}) in terms of the change in moles of \ce{Fe2O3}, 
\begin{equation}
    \Delta n_\mathrm{\ce{FeO}} = \left( \frac{N - n_\mathrm{\ce{Fe2O3}} + 2 \alpha\,n_\mathrm{\ce{FeO}}}{2 \alpha N - 2 \alpha\,n_\mathrm{\ce{FeO}} + n_\mathrm{\ce{Fe2O3}}} \right) \Delta n_\mathrm{\ce{Fe2O3}} \equiv \beta\,\Delta n_\mathrm{\ce{Fe2O3}} ~~, 
    \label{eqn:Total FeO}
\end{equation}
where $\alpha$ is the proportionality term from Equation \ref{eqn:Combined Param},
\begin{equation}
    \alpha = 1.504\,\frac{X_\mathrm{\ce{Fe^3+}}}{X_\mathrm{\ce{Fe^2+}}}~~.
\end{equation}
When reducing the melt phase, we thus choose the change in moles of \ce{Fe2O3}, $\Delta n_\mathrm{\ce{Fe2O3}}$, and Equation \ref{eqn:Total FeO} predicts the total change in \ce{FeO} across both reactions \ref{eqn:FMQ Reaction} \& \ref{eqn:IW Reaction}, which we repeat here,
\begin{equation}
    \ce{Fe_2O_3$^\text{melt}$ + H_2$^\text{gas}$ <=> 2FeO$^\text{melt}$ + H_2O$^\text{gas}$}~~,
    \label{eqn:FMQ Reaction Appendix}
\end{equation}
\begin{equation}
    \ce{FeO$^\text{melt}$ + H_2$^\text{gas}$ <=> Fe^$\text{metal}$ + H_2O$^\text{gas}$}~~,
    \label{eqn:IW Reaction Appendix}
\end{equation}
\begin{equation}
    \Delta n_\mathrm{\ce{FeO}}\,_{(tot)} = \Delta n_\mathrm{\ce{FeO}}\,_{(\ref{eqn:FMQ Reaction Appendix})} + \Delta n_\mathrm{\ce{FeO}}\,_{(\ref{eqn:IW Reaction Appendix})}~~.
\end{equation}
The change in moles of \ce{H2} and \ce{H2O} in the atmosphere, and \ce{Fe} in the metal phase, are then predicted via
\begin{equation}
    \begin{split}
        \Delta n_\mathrm{\ce{Fe}} &= - \Delta n_\mathrm{\ce{FeO}}\,_{(\ref{eqn:IW Reaction Appendix})} \\
        &= - \left(\Delta n_\mathrm{\ce{FeO}}\,_{(tot)} - \Delta n_\mathrm{\ce{FeO}}\,_{(\ref{eqn:FMQ Reaction Appendix})} \right) \\
        &= -(2 + \beta)\,\Delta n_\mathrm{\ce{Fe2O3}}~~,
    \end{split}
    \label{eqn:Fe reduce appendix}
\end{equation}
\begin{equation}
    \Delta n_\mathrm{\ce{H2}} = \Delta n_\mathrm{\ce{Fe2O3}} - \Delta n_\mathrm{\ce{Fe}} = (3 + \beta)\,\Delta n_\mathrm{\ce{Fe2O3}}~~,
    \label{eqn:H2 reduce appendix}
\end{equation}
\begin{equation}
    \Delta n_\mathrm{\ce{H2O}} = - \Delta n_\mathrm{\ce{H2}} = -(3 + \beta)\,\Delta n_\mathrm{\ce{Fe2O3}}~~.
    \label{eqn:H2O reduce appendix}
\end{equation}
When oxidising the melt phase, we instead choose the change in moles of \ce{Fe}. Equations \ref{eqn:Total FeO} \& \ref{eqn:Fe reduce appendix} yield the change in \ce{Fe2O3}, \ce{FeO}, \ce{H2}, and \ce{H2O} as a function of $\Delta n_\mathrm{\ce{Fe}}$, 
\begin{equation}
    \Delta n_\mathrm{\ce{Fe2O3}} = - \left(\frac{1}{2 + \beta}\right) \,\Delta n_\mathrm{\ce{Fe}}~~,
    \label{eqn:Fe2O3 oxidise appendix}
\end{equation}
\begin{equation}
    \Delta n_\mathrm{\ce{FeO}} = - \left(\frac{\beta}{2 + \beta}\right) \,\Delta n_\mathrm{\ce{Fe}}~~,
    \label{eqn:FeO oxidise appendix}
\end{equation}
\begin{equation}
    \Delta n_\mathrm{\ce{H2}} = - \left(\frac{3 + \beta}{2 + \beta}\right) \,\Delta n_\mathrm{\ce{Fe}}~~,
    \label{eqn:H2 oxidise appendix}
\end{equation}
\begin{equation}
    \Delta n_\mathrm{\ce{H2O}} = \left(\frac{3 + \beta}{2 + \beta}\right) \,\Delta n_\mathrm{\ce{Fe}}~~.
    \label{eqn:H2O oxidise appendix}
\end{equation}

\subsection{Simultaneous oxidation or reduction of a basaltic melt}
The prescription of \citet{kress1991compressibility} can be written as,
\begin{equation}
    \ln\left(\frac{X_\mathrm{\ce{Fe2O3}}}{X_\mathrm{\ce{FeO}}}\right) = a\ln\left( f\mathrm{O}_2 \right) - 1.828 \left(\frac{X_\mathrm{\ce{FeO}} + 2X_\mathrm{\ce{Fe2O3}}}{1 + X_\mathrm{\ce{Fe2O3}}} \right) + \sum_{i \neq \mathrm{FeO*}} \frac{d_i X_i\prime}{1 + X_\mathrm{\ce{Fe2O3}}} + \gamma(P,T) ~~,
    \label{eqn:Connecting KC}
\end{equation}
where 1.828 derives from the $d_\mathrm{FeO*}$ coefficient of Table \ref{tab:Kress and Carmichael}, and $\gamma(P,T)$ are the terms from Equation \ref{eqn:Kress and Carmichael} that are independent of composition. Because FeO* represents an assumption that all iron in the sample is in the form of \ce{FeO}, we must account for the presence of \ce{Fe2O3} ourselves through splitting up the FeO* term. The $(1 + X_\mathrm{\ce{Fe2O3}})$ terms re-normalise the sum of the mole fractions to unity. 

Next, from the Fe-FeO equilibrium, we can express the equilibrium constant for reaction \ref{eqn:IW Reaction Appendix} as, 
\begin{equation}
    k_\text{eq}^{(\ref{eqn:IW Reaction Appendix})} = \frac{a_{\ce{Fe}}~(f\ce{O2})^{1/2}}{a_{\ce{FeO}}} \approx \frac{0.98~ (f\ce{O2})^{1/2}}{X_{\ce{FeO}}}~~,
    \label{eqn:IW Equilibrium Q}
\end{equation}
where $0.98$ is the iron purity of the metal phase, and we have approximated a pure silicate phase such that the activity of each species is equal to its molar fraction (i.e., $a_\mathrm{FeO} = \gamma_\mathrm{FeO}X_\mathrm{FeO} \approx X_\mathrm{FeO}$). We take $k_\text{eq}^{(\ref{eqn:IW Reaction Appendix})}$ from \citet{frost1991introduction}, 
\begin{equation}
    \log_{10}(k_\text{eq}^{(\ref{eqn:IW Reaction Appendix})}) = -\frac{2748}{T} + 6.702 + 0.055\left(\frac{P - 1}{T}\right)~~.
    \label{eqn:Frost No Adjust}
\end{equation}
Substituting into Equation \ref{eqn:Connecting KC}, we obtain 
\begin{equation}
    \ln\left(\dfrac{X_\mathrm{\ce{Fe2O3}}}{X_\mathrm{\ce{FeO}}}\right) = 2a\ln\left(\dfrac{k_\text{eq}^{(\ref{eqn:IW Reaction Appendix})} X_\mathrm{FeO}}{0.98}\right)  - 1.828 \left(\dfrac{X_\mathrm{\ce{FeO}} + 2X_\mathrm{\ce{Fe2O3}}}{1 + X_\mathrm{\ce{Fe2O3}}} \right) + \sum_{i \neq \mathrm{FeO*}} \dfrac{d_i X_i\prime}{1 + X_\mathrm{\ce{Fe2O3}}} + \gamma(P,T) ~~.
    \label{eqn:Connecting Combined}
\end{equation}
We then express the change in $X_{\ce{FeO}}$ in terms of the change in $X_{\ce{Fe2O3}}$, 
\begin{equation}
    \Delta X_\mathrm{FeO} = \left( \dfrac{\dfrac{X_\mathrm{FeO} (1 + X_\mathrm{Fe2O3})}{X_\mathrm{Fe2O3}} + 1.828 X_\mathrm{FeO} \left( 2 - \dfrac{X_\mathrm{FeO} + 2 X_\mathrm{Fe2O3}}{1 + X_\mathrm{Fe2O3}} \right) + \dfrac{X_\mathrm{FeO}}{1 + X_\mathrm{Fe2O3}} \sum_{i=\mathrm{FeO*}} d_i X_i\prime}{(2a + 1)(1 + X_\mathrm{Fe2O3}) - 1.828 X_\mathrm{FeO}} \right) \Delta X_\mathrm{Fe2O3}
    \label{eqn:epsilon definition}
\end{equation}
Similarly to the peridotitic melt calculations, we convert from changes in molar fraction to changes in moles using Equations \ref{eqn:Delta X_fe_3} \& \ref{eqn:Delta X_fe_2}. For Equation \ref{eqn:Delta X_fe_3}, however, we must remove the factor of $2$ to account for the fact that we have $X_\mathrm{\ce{Fe2O3}}$ rather than $X_\mathrm{\ce{Fe^3+}}$. We yield the expression,
\begin{equation}
    \Delta n_\mathrm{\ce{FeO}} = \left( \frac{N - n_\mathrm{\ce{Fe2O3}} + \epsilon\,n_\mathrm{\ce{FeO}}}{\epsilon N - \epsilon\,n_\mathrm{\ce{FeO}} + n_\mathrm{\ce{Fe2O3}}} \right) \Delta n_\mathrm{\ce{Fe2O3}} \equiv \zeta\,\Delta n_\mathrm{\ce{Fe2O3}} ~~, 
    \label{eqn:Molar Relation}
\end{equation}
where $\epsilon$ is the proportionality term from Equation \ref{eqn:epsilon definition}. Equations \ref{eqn:Fe reduce appendix}-\ref{eqn:H2O oxidise appendix} are then used to calculated the changes in moles of system species, using $\zeta$ instead of $\beta$.

%% file: appendices/C_post_impact.tex
\newpage\section{The post-impact atmosphere}\label{sec:Appendix post-impact}
\renewcommand{\theequation}{C\arabic{equation}}
\renewcommand{\thetable}{C\arabic{table}}
\begin{figure}[h!]
    \centering
    \includegraphics[width=0.8\textwidth]{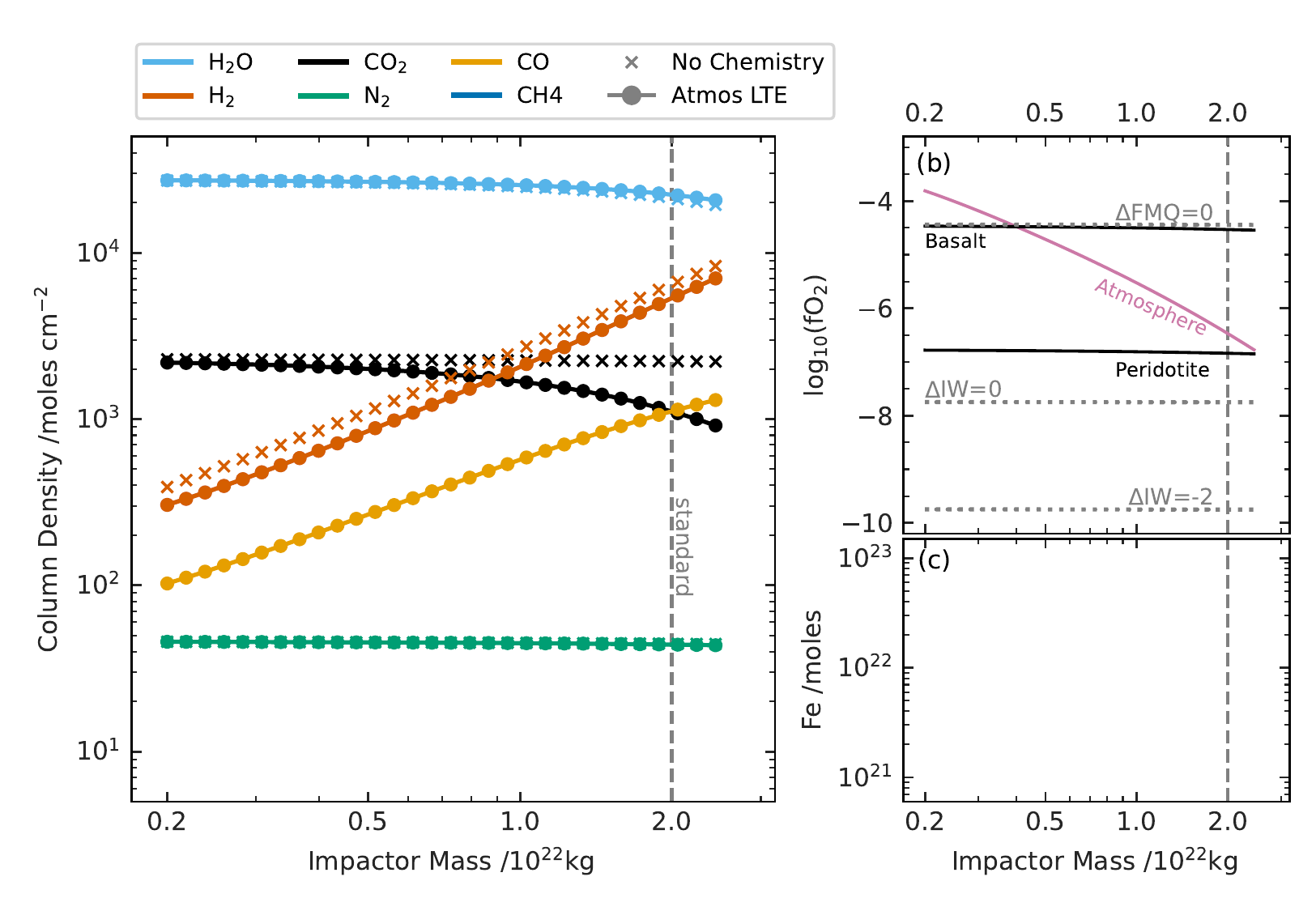}
    \caption{The state of the system at the post-impact stage as a function of impactor mass in Model 3B. All model parameters except for the varying impactor mass are the standard values detailed in Table 1. (a) Composition of the atmosphere with (circles) and without (crosses) the final processing step of local thermochemical equilibrium (LTE). (b) Oxygen fugacities of the atmosphere, basaltic melt and peridotitic melt. The FMQ and iron w\"{u}stite (IW) buffers are shown for reference (dotted lines). (c) Metallic iron remaining in the melt pool after metal saturation has been reached for each melt phase composition.}
    \label{fig:Appendix 6B}
\end{figure}
\begin{figure}[h!]
    \centering
    \includegraphics[width=0.9\textwidth]{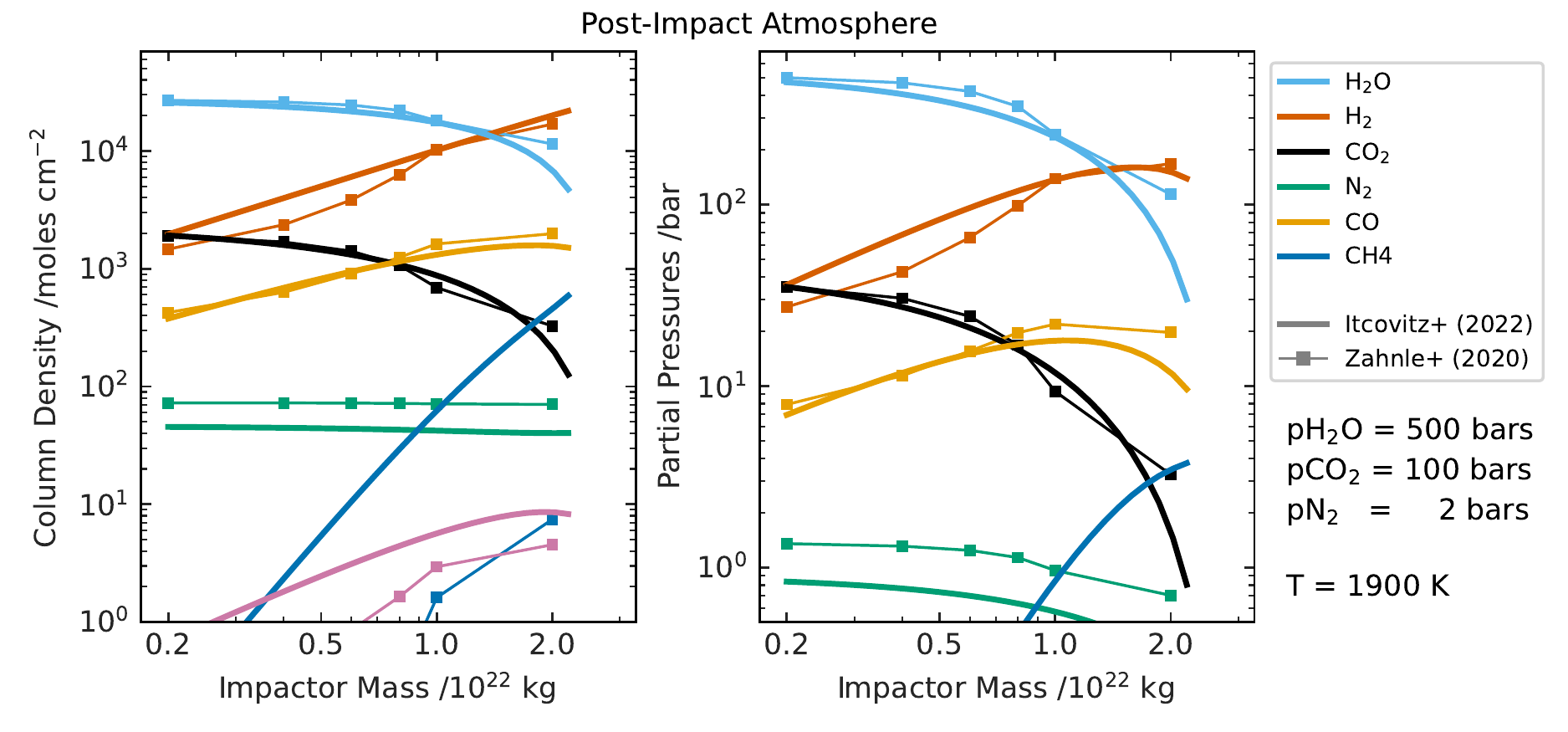}
    \caption{Comparison of the post-impact atmosphere of this study with the composition of the atmosphere of \citet{zahnle2020creation} once cooled to $1900\,\mathrm{K}$.}
    \label{fig:Appendix Zahnle}
\end{figure}